\documentclass[aps,preprint,nofootinbib]{revtex4}
\usepackage{color}
\usepackage{graphicx}
\usepackage{feynmp}
\usepackage{amsmath,amssymb}
\usepackage{url}
\usepackage{subfigure}
\newcommand{\lsim}{\mathrel{\mathop{\kern 0pt \rlap
  {\raise.2ex\hbox{$<$}}}
  \lower.9ex\hbox{\kern-.190em $\sim$}}}
\newcommand{\gsim}{\mathrel{\mathop{\kern 0pt \rlap
  {\raise.2ex\hbox{$>$}}}
  \lower.9ex\hbox{\kern-.190em $\sim$}}}
\newcommand{\gev}{{\,{\rm GeV}}}
\newcommand{\tev}{{\,{\rm TeV}}}
\newcommand{\al}{{\alpha}}

\newcommand{\lm}{{\lambda}}
\newcommand{\Lm}{{\Lambda}}
\newcommand{\gm}{{\gamma}}

\newcommand{\Gm}{{\Gamma}}

\newcommand{\Th}{{\Theta}}
\newcommand{\rd}{{\partial}}

\newcommand{\beq}{\begin{equation}}
\newcommand{\eeq}{\end{equation}}
\newcommand{\bea}     {\begin{eqnarray}}
\newcommand{\eea}     {\end{eqnarray}}
\newcommand{\no}{{\nonumber}}

\newcommand{\ptmiss}{{\rlap/p_T}}

\newcommand{\M}{{\mathcal{M}}}
\newcommand{\R}{{\mathcal{R}}}

\newcommand{\on}{ {(1)} }

\newcommand{\zt}{ { Z^{(2)} }}
\newcommand{\lo}{ { L^{(1)} }}

\newcommand{\neu}{\tilde{\chi}^0}

\newcommand{\neuo}{{\tilde{\chi}^0_1}}
\newcommand{\neut}{{\tilde{\chi}^0_2}}

\newcommand{\achc} {  \cosh^{-1}\chi }

\newcommand{\shz} {  s_{\eta_a} }
\newcommand{\chz} {  c_{\eta_a} }

\newcommand{\chet} {  c_{2\eta_B} }

\newcommand{\mmin} { m_{\rm min} }
\newcommand{\mmax} { m_{\rm max} }
\newcommand{\mcusp} { m_{\rm cusp} }
\newcommand{\mknee} { m_{\rm knee} }

\newcommand{\dphif}{ {d \widehat{\Phi}_4} }
\newcommand{\phif}{ { \widehat{\Phi}_4} }

\begin{document}

\preprint{PITT-PACC 1203}

\title{Kinematic Cusps With Two Missing Particles {\rm I}: \\  
Antler Decay Topology}
\author{Tao Han$^{1}$, Ian-Woo Kim$^2$,  Jeonghyeon Song$^{3}$}
\affiliation{
$^1$  Pittsburgh Particle physics, Astrophysics, and Cosmology Center, Department of Physics $\&$ Astronomy, University of Pittsburgh, 3941 O'Hara St., Pittsburgh, PA 15260, USA\\
$^2$Department of Physics, University of Michigan, USA\\
$^3$Division of Quantum Phases \& Devises, School of Physics, 
Konkuk University,
Seoul 143-701, Korea  }
\begin{abstract}
The kinematics of a final state system with two invisible particles and two visible particles {can}
develop cusped peak structures. 
This happens when the system has a fixed invariant mass (such as from a narrow resonant
particle decay or with a fixed collision c.m.~energy) and undergoes decays of two on-shell intermediate particles.
Focusing on the ``antler decay topology'', we derive general  analytic expressions for the invariant mass 
distribution  and the kinematic cusp position. The sharp cusp peaks and the endpoint positions can help to determine the masses of the missing particles and the intermediate particles. We also consider transverse 
momentum variables and angular variables. 
In various distributions the kinematic cusp peaks are present and pronounced. 
We also study the effects on such kinematic cusp structures 
 from realistic considerations including finite decay widths, the longitudinal boost of the system, and 
spin correlations. 
\end{abstract}

\maketitle

\section{Introduction}
\label{sec:introduction}

With the advent of the Large Hadron Collider (LHC), 
the TeV scale physics will be fully explored in the coming decades. 
Most pressing of all to learn is the mechanism of the electroweak symmetry breaking and 
the related underlying dynamics beyond the standard model (SM). 
Among many interesting phenomena associated with the new physics at the TeV
scale, the signature of events with large missing energy 
is one of the most exciting possibilities at the LHC. 
This is expected from new particles that do not leave any trace in 
the hadronic and electromagnetic components of the detector.
These new missing particles may help to address one of 
the most profound puzzles in cosmology: what constitutes nearly a quarter of the 
energy density of our current universe in a form of cold dark matter (CDM)~\cite{WMAP}. 
The thermal history of the early 
universe suggests that a stable neutral particle of the electroweak-scale mass and interaction, 
called the Weakly Interacting Massive 
Particles {(WIMP)}, is a plausible explanation of CDM~\cite{WIMP}
and may be discovered as a missing particle at TeV-scale colliders.

Missing energy signal is generic in many new physics models.
Additional discrete symmetry is often introduced to prohibit dangerous processes 
such as proton decay and to make the model compatible with the electroweak precision tests. 
Such a discrete symmetry (or parity)
often needs nontrivial representations of new particles, 
while it assigns vanishing charges (or trivial representation)  to the SM particles. 
Therefore, the lightest new particle is stable,
becoming a natural candidate for the CDM particle.
One of the most studied examples is the lightest neutralino 
in supersymmetric (SUSY) theories 
with $R$-parity conservation~\cite{SUSY:DM}.
Other examples include the lightest Kaluza-Klein (KK) particle 
in universal extra dimensional (UED) theories 
with KK parity conservation~\cite{DMExtraD}, 
and the heavy photon 
in the little Higgs models 
with $T$-parity conservation (LHT)~\cite{DMLittleHiggs}. 
In this regard, the search for missing particles 
at the LHC and future colliders 
has great implications in understanding 
both the fundamental particle physics
and the nature of our universe. 
 At hadron colliders,   the experimentally observable signature will be 
missing energy-momentum {\em transverse} to the beam direction.
Great efforts have been made on the phenomenological studies 
of the missing energy
signals in various new physics models \cite{Baer:2008uu,ETmiss} and optimistic
conclusions have been reached such that significant excess is expected 
above the SM background at the LHC \cite{TDRs}. 

In order to reveal the CDM identity and to compare with the results from direct and indirect DM searches, 
it is imperative to determine the missing particle  mass at colliders. 
 This is  a very challenging task  since such weakly  interacting neutral particles
leave neither charged particle tracks nor significant energy deposit 
in the detector.
Furthermore, the missing particles always come 
in pairs in an event due to the
conserved ``parity'',  so that the final state kinematics is under-constrained.
Finally, if we consider hadronic collisions such as at the LHC,
the partonic c.m.~energy as well as the frame are unknown on an even-by-event basis.

As reviewed in Ref.~\cite{Burns:Review},
most of the techniques for the missing particle mass measurement
at the LHC can be categorized into the following three cases:
(i) endpoint methods~\cite{endpoint};
(ii) polynomial methods~\cite{polynomial:combine,polynomial};
(iii) $M_{T2}$ methods~\cite{MT2:original,MT2,MT2:kink,subsystem:MT2}.
All three methods rely on a cascade decay of a heavy
new particle, ended up with a single missing particle $X$.
At each step of a cascade chain, a visible particle
is produced,
which may provide information on 
the missing particle mass as well as the intermediate new
particle mass.

Endpoint methods use the kinematic edges of 
invariant mass distributions of the visible
particles in a given cascade decay.
If the cascade chain is long enough 
with at least three visible particles,
the number of kinematical constraints 
is sufficient to determine all the masses involved~\cite{Burns:Review}.
When the decay chain is not long,  the observables are insufficient 
for complete mass determination. 
In addition, the positions of endpoints are more 
sensitive to the mass difference than to the absolute mass.

Polynomial methods use reconstructable events in which 
the number of the on-shell kinematic constraints exceeds the number of 
the unknown masses and momentum components. By combining multiple event 
information, one can maximize the information 
for determination of mass parameters~\cite{polynomial:combine}. 
However, this method intrinsically requires a long decay chain,
at least 
two-step cascade decays in each chain, producing four visible 
particles~\cite{Burns:Review}.
It suffers from small statistics and large combinatoric 
background. 

The $M_{T2}$ variable, originally proposed in Ref.~\cite{MT2:original},
is useful at hadron colliders for measuring the mass of a new mother particle
when pair-produced.
Two mother particles decay through the same decay chain.
For each chain, the transverse mass is constructed
with the missing transverse momentum.
As a function of a trial mass for the missing particle, 
$M_{T2}$ is the minimum value of the larger
value of these two transverse masses.
The minimization is over all possible missing transverse
momenta of two decay chains 
as satisfying the observed total missing energy constraint.
The $M_{T2}$ distribution has the maximum at the mother particle
mass when the trial mass hits the true missing particle mass. 
Therefore, it provides one relation between the mother particle mass and 
the missing particle mass. 
A more exciting observation is that the endpoint curve of $M_{T2}$
as a function of the trial mass
shows a kink where the trial mass becomes the true mass~\cite{MT2:kink}.

In all three methods above,
a crucial issue is how to fully reconstruct the kinematics of a signal event. 
This relies on exclusive selection of events  of a given type.
If the decay chain is long,
the reconstruction becomes more difficult as combinatoric complications emerge:
the large number of involved particles entangle the origin of the decay of 
each observed particle.
The hemisphere method, 
an algorithm to group collinear and high-$p_T$
particles, 
was shown to be useful to some extent in the inclusive $M_{T2}$
analysis for the disentanglement
of the data~\cite{Nojiri:hemisphere}.

\begin{figure}[!t]
\centering
\includegraphics[scale=1]{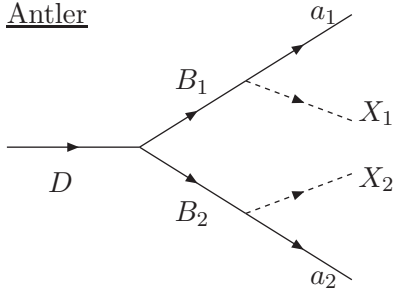}
  \caption{\label{fig:feyn:antler}
The antler decay topology of a heavy new particle $D$ into two missing particles
($X_1$ and $X_2$) and two visible 
particles ($a_1$ and $a_2$).}
\end{figure}

\begin{figure}[!t]
\centering
\includegraphics[scale=1]{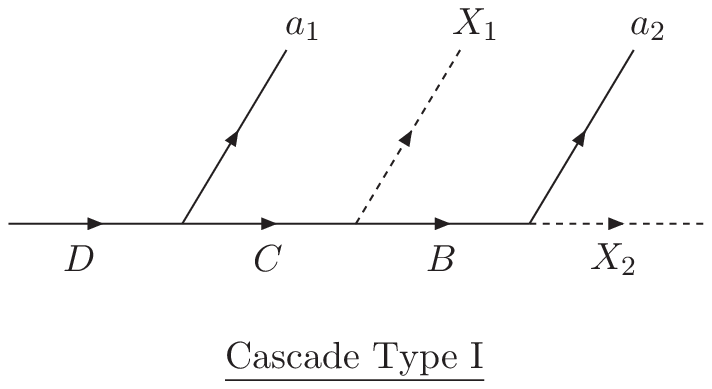}
\phantom{xxx}
\includegraphics[scale=1]{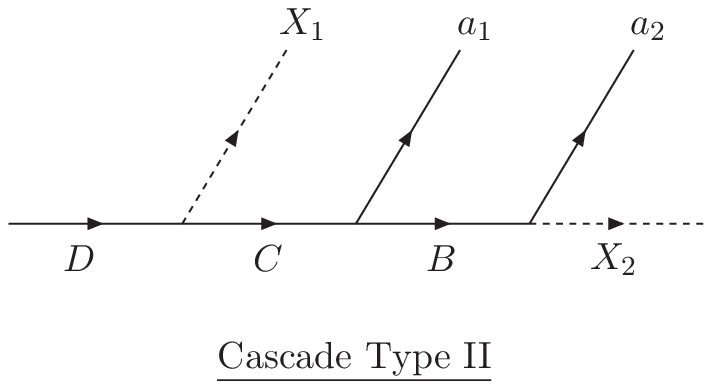}
  \caption{\label{fig:feyn:cas}
The cascade decay topology of a heavy new particle $D$ into two missing particles
($X_1$ and $X_2$) and two visible 
particles ($a_1$ and $a_2$).}
\end{figure}

Recently, it has been pointed out that the missing particle 
mass can be determined from singular structures in kinematic
distributions for shorter and simpler decay chains~\cite{cusp,Kim:2009si}. 
In our previous work~\cite{cusp}, we considered 
a resonant ``antler decay''  of a heavy new particle into a pair of missing particles
and a pair of SM visible particles, as shown in Fig.~\ref{fig:feyn:antler}, and found
non-smooth peaks in some kinematic distributions.
These peaks are called ``cusps''  and the
positions of the cusps depend only on the masses 
of the involved particles.
The cusp is statistically more advantageous
because it is 
at the peak region.  
The mass measurements can be benefited
from knowing the kinematic cusp structures. 

We consider a resonant decay of a heavy particle $D$ into
two visible particles and two missing particles.
The invariant mass distributions of this type of decay were first presented 
in our previous publication~\cite{cusp}, 
and recently further studied in Ref.~\cite{Invariant:mass}.
Obviously this heavy particle $D$ is parity-even.
The general topology of such resonance decays 
is divided into two classes:
\begin{enumerate}
\item
{\em Antler decay topology: }  
a heavy particle $D$
decays into two parity-odd particles ($B_1$ and $B_2$) at the first step 
and each parity-odd particle subsequently decays into a missing (denoted by dashed lines)
particle and a visible particle,
as in Fig.~\ref{fig:feyn:antler}.
\item
{\em Cascade decay topology:}   
a heavy particle $D$ splits to two particles with one or both visible particles
at each step,
finally into a missing particle. 
According to at which step the first missing particle comes out,
there are two non-trivial cascade topologies, as in Fig.~\ref{fig:feyn:cas}.
\end{enumerate}

The antler decay
and the cascade decay are siblings to each other 
as they share the same skeleton of topology.
Since they have different orientation of 
incoming and outgoing particles,
the cusps appear with different manifestations.
In this paper, we focus on the antler topology only and 
leave the presentation on the cascade decay topology
to a companion paper~\cite{long:cascade}.

Antler decays arise in many new physics models.
We now list a few examples for illustration.
\begin{itemize}
\item 
In the Minimal Supersymmetric Standard Model (MSSM), 
the heavy  CP-even neutral Higgs bosons may have sizable rates of
the following decay~\cite{Higgs:MSSM}:
\begin{eqnarray}
H \rightarrow \neut + \neut \rightarrow Z \neuo + Z \neuo.
\label{eq:mssm}
\end{eqnarray}
\item In the MSSM with an additional $U(1)$ gauge interaction, 
the extra $U(1)$ gauge boson $Z'$ can have antler decay modes like \cite{Zp}
\begin{eqnarray}
Z' \rightarrow \tilde \ell^- + \tilde \ell^+ \rightarrow \ell^- \neuo + \ell^+ \neuo.
\label{eq:zp:susy}
\end{eqnarray}
\item 
The ultraviolet completion of the LHT model 
often involves an
extension of the Higgs sector that accommodates 
heavy Higgs bosons.
Large top Yukawa coupling
leads to substantial decay of the neutral heavy Higgs
into a pair of $T$-parity odd top quarks $t_-$,
followed by $t_-$ decay into the SM top quark
and the heavy photon 
$A_H$ (the CDM candidate)~\cite{DMLittleHiggs}:
\begin{eqnarray}
H \rightarrow t_- +  \bar{t}_- \rightarrow t A_H  + \bar{t} A_H. 
\label{eq:lht}
\end{eqnarray}
\item 
In the UED model with KK parity conservation, 
the second KK mode
of the $Z$ boson can have antler decay modes~\cite{KK2:mUED}.
$ Z^{(2)}$ decays into a pair of the first KK modes of 
the lepton, followed by its decay into a SM lepton and the
CDM particle $B^{(1)}$:
\begin{eqnarray}
Z^{(2)} \rightarrow L^{(1)} + L^{(1)} 
\rightarrow \ell^- B^{(1)} 
+ \ell^+ B^{(1)}. 
\label{eq:mued}
\end{eqnarray}
\item At lepton colliders with $e^{+}e^{-}$ or  $\mu^{+}\mu^{-}$ collisions, 
the well-determined c.m.~energy renders 
some pair production and their subsequent decay
processes to be of the antler topology.
One example is
\begin{eqnarray}
\label{eq:ilc}
 e^{+}e^{-}/\mu^{+}\mu^{-} \to
\tilde \ell^{+} + \tilde \ell^{-} \to  \ell^+ \neuo + \ell^- \neuo .
\end{eqnarray}
\end{itemize}

In the current work, we only focus on the
generic features of antler kinematics. 
The rest of the paper is organized as follows.  
We begin our discussion by explaining the unique features of the antler kinematics in Sec.~\ref{sec:kinematics}.  
Focused on the symmetric antler decay,
we consider the massive visible particle case,
and demonstrate in Sec.~\ref{sec:massive}
the cusps and endpoints in the 
kinematic distributions of the invariant mass,
transverse momenta, and angular variables,
constructed from two visible particles.
In Sec.~\ref{sec:massless},
we study the massless visible particle case. 
We discuss some effects of more realistic considerations in Sec.~\ref{sec:real:effects}, such as the 
finite decay widths of the resonant particles, the longitudinal boost between 
the c.m.~frame and the lab frame, and spin correlations.
We conclude in Sec.~\ref{sec:conclusions}. 
A few appendices are devoted to some technical details
for a general four-body phase space treatment, 
the derivations of the cusp peak and analytic expressions
of some kinematic distribution,
and more discussions for the general antler decay.

\section{Kinematics of antler decay topology  with two missing particles}
\label{sec:kinematics}

We consider the resonant decay of a heavy particle $D$ into two visible 
particle $a_1$ and $a_2$, and 
two missing particle $X_1$ and $X_2$ via a chain of two-body decays 
through intermediate particles $B_1$ and $B_2$, as depicted in Fig.~\ref{fig:feyn:antler}:
\bea
\label{eq:antler}
D(P) & \to& B_1(p_1) +B_2(p_2),
\\ \no  &&
B_1(p_1) \to a_1(k_1)+ X_1(k_3), \quad 
B_2(p_2) \to a_2(k_2) +X_2(k_4).
\eea
Since most of the processes of our interest are symmetric between two decay branches,
we focus on the symmetric antler decay, defined by
\beq
\hbox{Symmetric antler: }
B \equiv B_1 = B_2,\quad a \equiv a_1 = a_2,\quad X \equiv X_1 = X_2
\label{eq:sym:antler}.
\eeq
The general antler decay with arbitrary masses  is to be discussed in
Appendix \ref{sec:general}. 

\begin{figure}[!t]
\centering
  \includegraphics[scale=0.2]{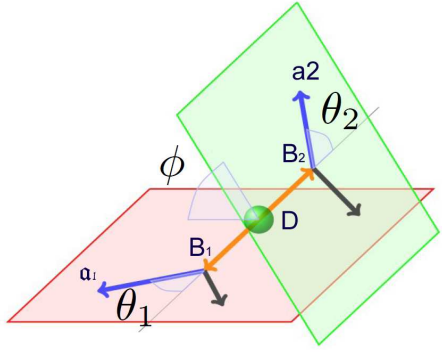}
  \caption{\label{fig:antphase} 
  Kinematic configuration of the antler decay in the rest frame of the 
  parent particle $D$ via two intermediate particles $B_{1}$ and $ B_{2}$,  
 followed by $B \to a X$. 
$\theta_1$ and $\theta_2$ are defined in the rest frames of $B_1$ and $B_2$, respectively, 
and $\phi$ in the $D$ rest frame. }
\end{figure}

In the three-dimensional momentum space, the kinematic configuration of the antler decay 
is illustrated in Fig.~\ref{fig:antphase}. 
In the rest frame of the parent particle $D$, 
the intermediate particles $B_1$ and $B_2$ are moving back-to-back, and the momentum 
direction defines the principal decay axis $z$, with $B_1$ moving into the $+z$ direction
and $B_{2}$  into the $-z$ direction. 
Two momenta of $a_1$ and $X_1$
in the $B_1$ rest frame form 
the decay plane $P1$, which is identified as the  $xz$-plane. 
In the same way, the decay plane $P2$ is defined by
the $B_2$ decay products. In the decay plane $P1$, 
we define a polar angle $\theta_1$ between the $+z$ direction and 
the $a_1$ momentum in the $B_1$ rest frame.  Similarly, 
$\theta_2$ is the polar angle 
between the $-z$ direction and the $a_2$ momentum in the $B_2$ rest frame. 
The azimuthal angle between two decay planes $P1$ and $P2$ 
is denoted by $\phi$. 

As explicitly shown in
Appendix \ref{sec:invmassderivation},
these three internal angles ($\theta_1$, $\theta_2$, and $\phi$)
represent the phase space configuration of the antler decay topology.
The dynamics of the antler decay 
is encoded in the differential decay width $d\Gamma$ 
defined in the rest frame of $D$. 
$d\Gamma$  is a function of the internal 
phase space variable $(\theta_1, \theta_2, \phi)$:
\beq
d \Gm  
\propto
\widehat{|\M|^2}\, \dphif ,
\eeq
where $\widehat{|\M|^2}$ is a reduced matrix elements and
$\dphif = d\cos \theta_1\, d \cos\theta_2 \, d\phi$
(see Appendix \ref{sec:invmassderivation} for more details).
The reduced matrix element $\widehat{|\M|^2}$ is a smooth function of 
$(\theta_1, \theta_2, \phi)$, and thus
$d\Gamma /  \dphif$  does not show any 
singular behavior.

Kinematic singularities emerge as missing particles allow us 
only the projection of the full kinematic
phase space onto a lower dimensional phase space accessible
by the visible particle momenta.
This partial access inevitably 
hides some of necessary information for the full mass 
reconstruction. 
However, we can still decode the mass information out of 
some observables, say $Y$'s.

In order to obtain $d\Gamma / dY$, 
we project the hypersurface of the phase space 
$(\theta_1, \theta_2, \phi)$ onto $Y$:
for each value of $Y$, $d \Gm /d Y$ is proportional to
the volume of the
hypersurface corresponding to that specific value of $Y$.
When the hypersurface fails to be a manifold at a certain
point $Y$, $d \Gm /d Y$ develops non-smoothness.
This is called 
singular points, where the differential  
 $(\partial Y / \partial \theta_1, \partial Y / \partial \theta_2, 
\partial Y / \partial \phi)$ vanishes~\footnote{In multi-dimensional cases, 
this condition is a reduced rank condition of Jacobian matrix of mapping from 
the phase space to the observable $Y$'s~\cite{Kim:2009si}.}.
As a result, we see non-smooth behaviors in the distribution 
of $Y$, which give rise to kinematic cusps and endpoints.  
General discussions on the development 
of singularity in the multi-dimensional observable phase space
have been presented in Ref.~\cite{Kim:2009si}.

Since the parent particle $D$ is moving in the lab frame,  
the observable variable $Y$ from
the momenta of two visible particles $a_{1}$ and $a_2$ 
can be classified into three categories:
\begin{itemize}
\item Lorentz-invariant observable:  there is only one Lorentz-invariant observable, 
the invariant mass of $a_1$ and $a_2$, 
    \bea
\label{eq:m:def}
    m = \sqrt{(k_1+k_2)^2}.
    \eea
\item Longitudinal-boost invariant observables: 
  \begin{itemize}
  \item the 
transverse momentum of one visible particle $i$:
    \bea
    p_{Ti} =\left| \mathbf{k}_{iT} \right| .
    \eea
Here and henceforth, a bold-faced letter denotes a three-momentum.
  \item the total transverse momentum of the $a_1$-$a_2$ system:
    \bea
	\label{eq:def:pT}
    p_{T} = \left| \mathbf{k}_{1T} + \mathbf{k}_{2T}\right|.
    \eea
    In the four-body decay under consideration, this
    is the same as the magnitude of the missing transverse momentum $\ptmiss$ 
    of the decay.  
    \item the transverse mass of the $a_1$-$a_2$ system:
    \bea
 m_T   = \sqrt{p_{T}^2 + m^2} .
    \eea
   \item the cluster transverse mass of the $a_1$-$a_2$-$\, \ptmiss$ system:
    \bea
    m_{T}^{C} = m_T  + \ptmiss. 
    \eea
   \item the rapidity difference:
    \beq
    \Delta \eta = |\eta_{a1} - \eta_{a2}|, \quad {\rm where}\ \  \eta_{ai} = {1 \over 2} \ln \left({ {E_{i} + k_{iz} } \over {E_{i}- k_{iz}} }\right).
    \eeq
  \end{itemize}
\item Non-invariant observable: we consider an angular variable $\Theta$, which is
 the angle between one visible particle (say $a_1$) 
in the  c.m.~frame of $a_1$ and $a_2$
and the c.m.~moving direction in the $D$ rest frame, given by 
    \bea
\label{eq:def:costh}
    \cos\Theta
    = - \frac{\mathbf{k}_1^{(a_1 a_2)} \cdot (\mathbf{k}_1
+\mathbf{k}_2 )^{(D)} }
    { \big|\mathbf{k}_1^{(a_1 a_2)} \big| ~
\big|  \mathbf{k}_1
+\mathbf{k}_2 \big|^{(D)} }.
    \eea
In what follows,
the superscript in a momentum denotes the 
reference frame.
In the main text, a momentum without a superscript is in the lab 
frame.
\end{itemize}

\section{Massive visible particle case}
\label{sec:massive}
In this section,
we consider the case of massive visible particles.
For a resonant decay,
it is very convenient to express the kinematics 
in terms of rapidity variables. 
For a two-body decay of $i \rightarrow j + k$,  
we write the four-momentum of the particle $j$ 
in the rest frame of the parent particle $i$
as $p_{j}^{(i)} = (E_{j}^{(i)}, \mathbf{p}_{j}^{(i)}) 
= (m_{j} \cosh\eta_{j},  m_{j}\sinh\eta_{j} \hat{\mathbf{p}}^{(i)}_j)$.
Here
$\eta_j$ is the rapidity of particle $j$ in the rest frame of the parent $i$,
given by 
\beq
\cosh \eta_{j} \equiv  {E_{j}^{(i)} \over m_{j}} = \frac{m_i^2 + m_j^2 - m_k^2 }{2 m_i m_{j}}. \\
\eeq
The superscript of a rapidity, specifying the reference frame, is omitted 
when it is the rest frame of the parent particle.
In this section, we assume that all the particles are massive.
The massless case will be covered in the next session 
by taking the massless limit from the massive case.

Now we illustrate the symmetric antler decay
defined in Eq.~(\ref{eq:sym:antler}),
which has two independent rapidity parameters
$\eta_B$ and $\eta_a$:
\begin{eqnarray}
\label{eq:eta:def}
\cosh \eta_B = \frac{m_D}{2 m_B}, \quad
\cosh \eta_a = \frac{m_B^2 - m_X^2 + m_a^2}{ 2 m_a m_B}. 
\end{eqnarray}
Note that $\eta_B$ 
is determined from 
$D \rightarrow B_1 B_2$ decay,
and  
$\eta_a $
from 
$B_1 \rightarrow a_1 X_1$ decay (or $B_2 \rightarrow a_2 X_2$ equivalently). 

In the $D$ rest frame, the momenta of the particles $a_1$ and $a_2$ are
\begin{eqnarray}
k^{(D)}_{1} & = &
m_a
\left( \begin{array}{c}
\cosh \eta_B \cosh \eta_{a} + \sinh \eta_B \sinh \eta_a \cos \theta_1\\
\sinh \eta_a \sin \theta_1 \\
0 \\
\sinh \eta_B \cosh \eta_a + \cosh \eta_B \sinh \eta_a  \cos \theta_1 
\end{array} \right), 
\label{eq:k1Drest} \\ 
k^{(D)}_{2} & = &
m_a
\left( \begin{array}{c}
\cosh \eta_B \cosh \eta_{a} + \sinh \eta_B \sinh \eta_a \cos \theta_2\\
\sinh \eta_a \sin \theta_2 \cos \phi\\
\sinh \eta_a \sin \theta_2 \sin \phi \\
-\sinh \eta_B \cosh \eta_a - \cosh \eta_B \sinh \eta_a  \cos \theta_2 
\end{array} \right),
\label{eq:k2Drest}
\end{eqnarray}
where the internal phase space angles of $(\theta_1,\theta_2,\phi)$ are defined
in Fig.~\ref{fig:antphase}.

\subsection{Invariant mass distribution}
\label{subsec:massive:m}
The invariant mass of the two visible particles $a_1$ and $a_2$ is explicitly
obtained from $k^{(D)}_{1}$ and $k^{(D)}_{2}$ in Eqs.~(\ref{eq:k1Drest})
and (\ref{eq:k2Drest}):
\begin{eqnarray}
m^2 &=& m_a^2 \bigg[
\left\{2 \cosh \eta_B \cosh \eta_a + \sinh \eta_B \sinh \eta_a 
(\cos \theta_1 + \cos \theta_2 ) \right\}^2  \nonumber \\
&& \qquad - \sinh^2 \eta_a (\sin \theta_1 + \sin \theta_2 \cos \phi)^2  
 -\sinh^2 \eta_a \sin^2 \theta_2 \sin^{2}\phi \nonumber\\
&& \qquad - \cosh^2 \eta_B \sinh^2 \eta_a (\cos \theta_1 - \cos \theta_2 )^2
\bigg] \, .
\label{eq:invmassintheta}
\end{eqnarray}

\begin{figure}[!t]
\centering
\subfigure[]{\includegraphics[height=37ex]{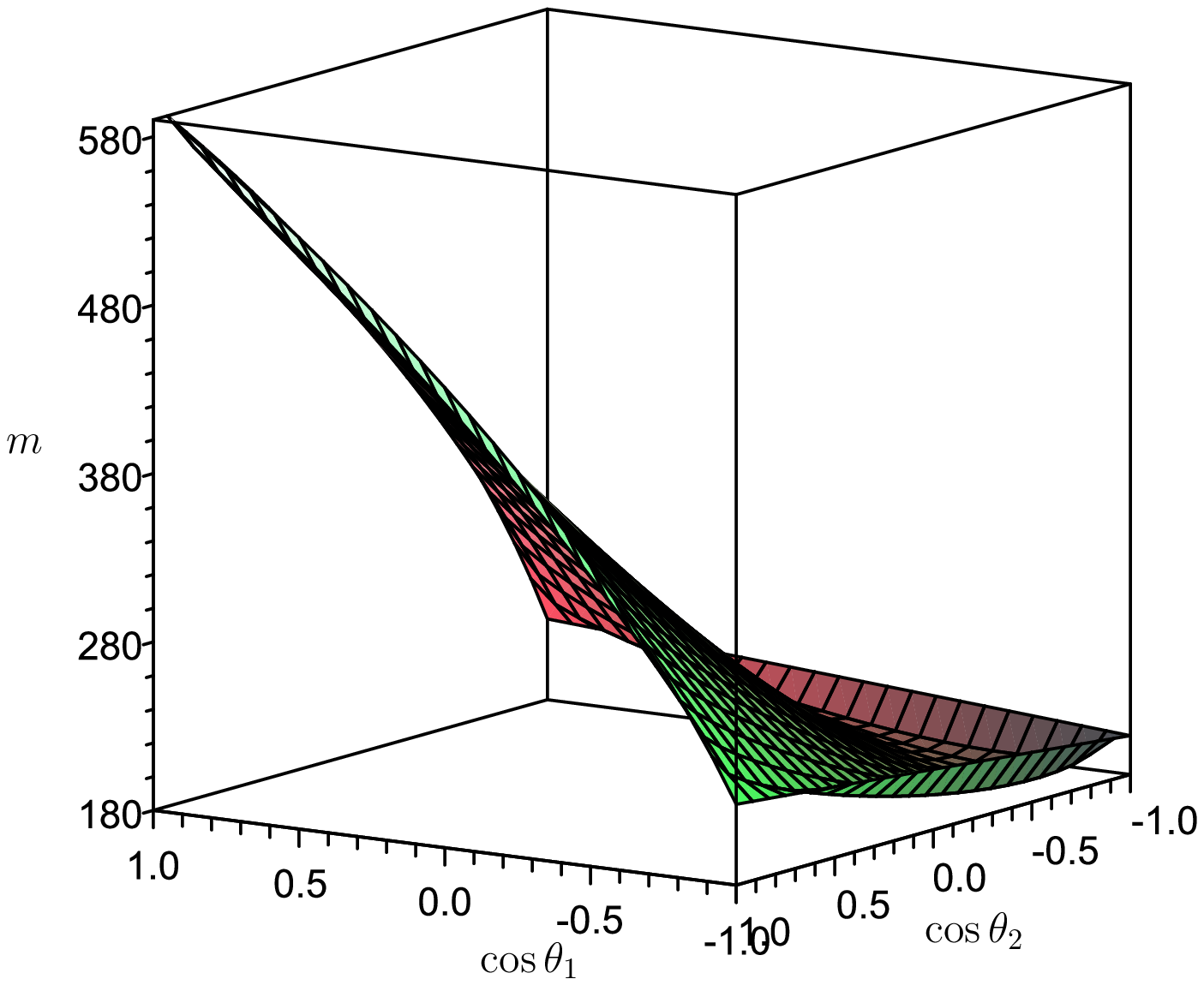}}
\subfigure[]{\includegraphics[height=37ex]{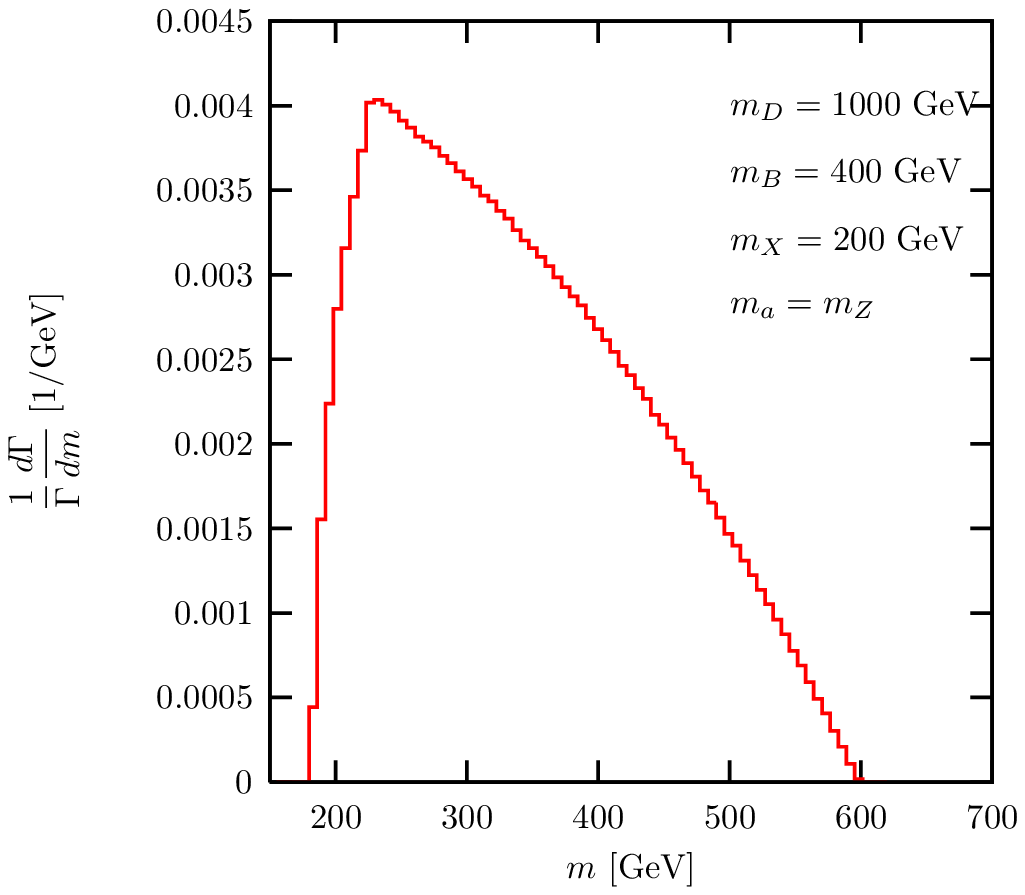}}
  \caption{\label{fig:massiveps} 
(a) The invariant mass $m$ 
as a function of $\cos \theta_1$ and $\cos \theta_2$, 
with $\phi = 0$.
(b) The normalized differential decay rate as a function of $m$. 
We take the mass parameters of
$m_D = 1\tev$, $m_B = 400\gev$,
$ m_X= 200\gev$ and $m_a = m_Z$.}
\end{figure}

In Fig.~\ref{fig:massiveps}(a), 
we show the invariant mass $m$ 
as a function of  $\cos \theta_1$ and 
$\cos \theta_2$. 
For the sake of illustration,  we take
$m_{D}=1\tev,\  m_{B}=400\gev,\  m_{a}= m_Z, \ m_X=200\gev$,  
and fixed $\phi = 0$.
The mapping of
this non-trivial hypersurface onto the $m$ 
yields a 
singular structure in the $d\Gamma/dm$ distribution as in Fig.~\ref{fig:massiveps}(b). 
To understand how this distinctive feature occurs, 
we study this mapping 
by examining the following some
critical points:
\begin{itemize}
\item \underline{\textsc{Point (i)}: $(\cos\theta_1, \cos\theta_2) = (1,1)$} \\
Since $a_1$ and $a_2$ move back-to-back in the $D$ rest frame,
their invariant mass becomes maximum.
The rapidity of $a_1$ in the rest frame of $D$ is
the same as that of $a_2$, such that
$|\eta_{a_1}^{(D)}| = |\eta_{a_2}^{(D)}| = \eta_B + \eta_a$. 
Therefore, the relative rapidity of $a_2$
with respect to $a_1$ is
$\eta^{(a_1)}_{a_2} = 2 (\eta_B + \eta_a)$. 
\item \underline{\textsc{Point (ii)}: $(\cos\theta_1, \cos\theta_2) = (\pm 1,\mp 1)$ }
\\
One visible particle, say $a_1$,
moves in the same direction of its parent $B_1$ with the rapidity of
$\eta^{(D)}_{a_1} = \eta_B + \eta_a$,
and the other visible particle $a_2$
moves in the opposite 
direction of its parent with
$\eta^{(D)}_{a_2} = | \eta_B - \eta_a |$. 
If $\eta_a > \eta_B$,  the directions of $a_1$ and $a_2$ 
in the $D$ rest frame are the same, which 
implies $\eta^{(a_1)}_{a_2} = \eta^{(D)}_{a_1} - \eta^{(D)}_{a_2}$.
If $\eta_a < \eta_B$,
the direction of $a_1$ and $a_2$ are opposite so that 
$\eta^{(a_1)}_{a_2} = \eta^{(D)}_{a_1} + \eta^{(D)}_{a_2}$. Therefore, regardless of the ordering of $\eta_a$ and 
$\eta_B$, $\eta^{(a_1)}_{a_2} = 2 \eta_B$.
Note that two configurations of $(\cos\theta_1, \cos\theta_2) = (1,-1)$ 
and $(\cos\theta_1, \cos\theta_2) = (-1, 1)$ 
are symmetric to each other.
\item \underline{\textsc{Point (iii)}: $(\cos\theta_1, \cos\theta_2) = (-1,-1)$} 
\\
$a_1$ and $a_2$ move in the opposite 
direction to $B_1$ and $B_2$
in their parent's rest frames, respectively. 
Their rapidities are $|\eta_{a_1}^{(D)}| = |\eta_{a_2}^{(D)}| =  | \eta_B - \eta_a |$,
leading to $\eta^{(a_1)}_{a_2} = 2 | \eta_B-\eta_a |$. 
\item \underline{\textsc{Point (iv)}:
$\theta_2 = \theta_1,\  \phi=0,\  \cos\theta_1 
= -\tanh \eta_B / \tanh \eta_a$
with  $\eta_a > \eta_B$}
\\
This special configuration gives rise to 
the same four-momenta of the two visible particles
as can be seen in Eqs.~(\ref{eq:k1Drest}) and (\ref{eq:k2Drest}).
$a_1$ and $a_2$ are relatively at rest, 
resulting in $\eta^{(a_1)}_{a_2} = 0$. 
The condition $\eta_a > \eta_B$ is required to
guarantee the equality of $k_1^{(D)}$ and $k_2^{(D)}$,
which cannot be achieved if the particle $B$ is boosted more highly than the particle $a$ (or equivalently $|\cos \theta_1| \leq 1$ for physical configurations).
\end{itemize}
\textsc{Point (i)} corresponds to the maximum endpoint, and 
\textsc{Points (ii)} to the cusped peak.
When \text{Point (iv)} exists, \textsc{Point (iii)} 
corresponds to the non-smooth kink, and
\textsc{Point (iv)} to the minimum endpoint at $m=2 m_a$.
Otherwise, \textsc{Point (iii)} becomes the
minimum endpoint at $m = 2 m_a \cosh (\eta_B - \eta_a)$.
 
\begin{figure}[!t]
\centering
\includegraphics[scale=0.6]{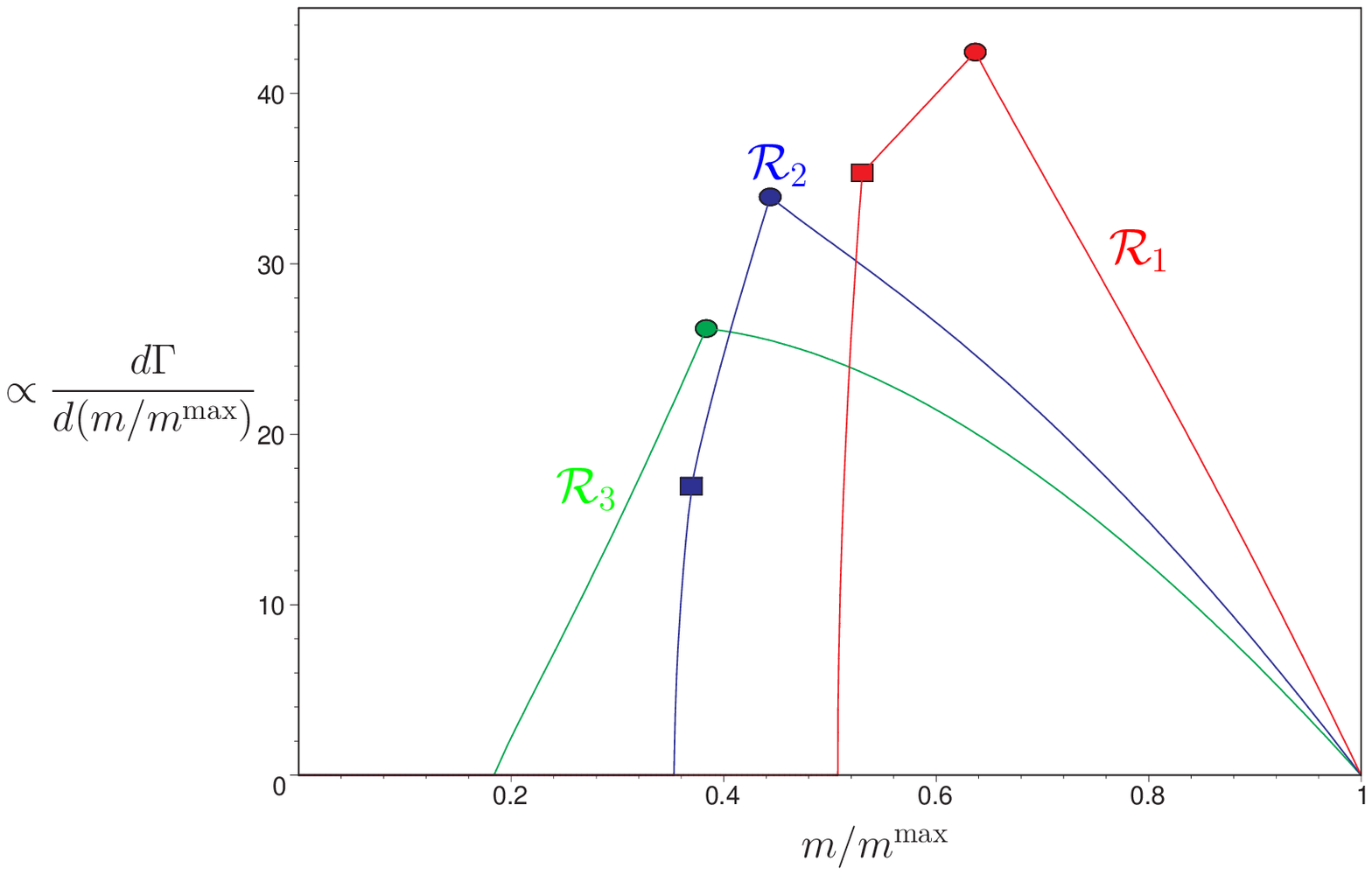}
  \caption{\label{fig:dgmdmhat}
The shapes of the function of $d\Gamma / d m$ 
for three representative regions,
$\R_1$, $\R_2$, and $\R_3$.
With the fixed ${\eta_a}=1$, we set $\eta_B=0.3$ 
for $\R_1$, $\eta_B =0.7$ for $\R_2$,
and $\eta_B=1.5$ for $\R_3$.
}
\end{figure}

Now we present the analytic expression
of the invariant mass distribution.
The functional forms are 
different in the following three mass regions:
\bea
\label{eq:R:explanation}
\R_1: \eta_B < \frac{{\eta_a}}{2},
\qquad
\R_2: \frac{{\eta_a}}{2} < \eta_B <{\eta_a},
\qquad
\R_3 :  {\eta_a}<\eta_B.
\eea

In Fig. \ref{fig:dgmdmhat}, we show the invariant mass distribution
$d\Gamma / d(m/\mmax)$ for $\R_1$, $\R_2$, and $\R_3$. 
Regardless of the parameter regions, the maximum 
endpoint of the $m$ distribution corresponds to \textsc{Point (i)}:
\begin{eqnarray}
\label{eq:mmax}
m_{\rm max} = 2 m_a \cosh (\eta_B+\eta_a). 
\end{eqnarray}  
For $\R_1$ and $\R_2$, the minimum endpoint occurs at
$m = 2 m_a$ while for $\R_3$ the minimum 
is different:
\begin{eqnarray}
m_{\rm min} = 
\begin{cases}
\begin{array}{ll} 
2 m_a, & \mbox{ for $\R_1$ and $\R_2$}, \\
2 m_a \cosh (\eta_B - \eta_a), & \mbox{ for $\R_3$}.
\end{array}
\end{cases}
\end{eqnarray} 
The condition of $\eta_B > \eta_a$ in $\R_3$
does not allow the equality of $k_1^{(D)} = k_2^{(D)}$
which would lead to $m_{\rm min} =  2 m_a$.
In $\R_1$ and $\R_2$, there are two non-smooth points in the middle of the distribution. 
Let us call the point at the smaller value of $m$ 
(marked by squares)
a knee point and the other point at the larger value of $m$
(marked by circles) a cusp point. 
In $\R_1$, the knee point corresponds to \textsc{Point (ii)} and
the cusp point to \textsc{Point (iii)}. 
In $\R_2$, it is opposite. 
In $\R_3$, there is only one sharp peak, the cusp.
We summarize the results of the minimum, cusp, knee, and maximum of
the $m$ distribution in Table \ref{table:m:massive}.

\begin{table}[!t]
\centering
\begin{tabular}{clll}
\hline
\hline
   & $~~~\R_1$ 
   & $~~~\R_2$ 
   & $~~~\R_3$ 
    \\
\hline 
$m_{\rm min}$~~~  &  ~~~$2 m_a$ & ~~~$2 m_a$ & ~~~$2 m_a{c}_{\eta_B-{\eta_a}}$ \\
$m_{\rm knee}$~~~ &  ~~~$2 m_a c_{\eta_B}$ &  ~~~$2 m_a c_{\eta_B -\eta_a}$ & ~~~~~~ -        \\
$m_{\rm cusp}$~~~ & ~~~$2 m_a{c}_{\eta_B-{\eta_a}}$ & 
~~~$2 m_a{c}_{\eta_B}$ & 
~~~$2 m_a{c}_{\eta_B} $\\
$m_{\rm max}$~~~ & ~~~$2 m_a c_{\eta_B + \eta_a}$ &
 ~~~$2 m_a c_{\eta_B + \eta_a}$ & 
 ~~~$2 m_a c_{\eta_B + \eta_a}$ \\
\hline
\hline\end{tabular}
\caption{\label{table:m:massive} Summary of the
minimum, cusp, knee, and maximum of
the $m$ distribution for the mass parameter regions $\R_1$, $\R_2$,
and $R_3$.
We have used a concise notation of $\cosh x \equiv c_x$. }
\end{table}

The invariant mass distributions for three mass regions are
\bea
\label{eq:dphi4:dm:R1}
\left. \frac{d \Gamma}{d m}\right|_{\R_1}
\!\!\!&\propto&
\left\{
   \begin{array}{ll}
   2 m\cosh^{-1}\left(\frac{m^2}{2 m_a^2}-1\right) ,
               & \hbox{ if }~ 2 m_a<m <2 m_a c_{\eta_B}; \\
   4 \eta_B \,m  ,
               & \hbox{ if }~ 2 m_a c_{\eta_B}< m <
2 m_a c_{{\eta_B}-{\eta_a}}; \\
   m\left[2(\eta_B+{\eta_a}) -\cosh^{-1}\left(\frac{m^2}{2 m_a^2}-1\right) \right] ,
               & \hbox{ if }~
2 m_a c_{{\eta_B}-{\eta_a}}<m 2 m_a c_{{\eta_B}+{\eta_a}};
\\
   0, & \hbox{ otherwise },
   \end{array}
 \right.
\eea
\bea
\left. \frac{d \Gamma}{d m} \right|_{\R_2}
\!\!\!&\propto&
\left\{
   \begin{array}{ll}
   2m\cosh^{-1}\left(\frac{m^2}{2 m_a^2}-1\right) ,
               & \hbox{ if }~2 m_a< m 2 m_a c_{{\eta_B}-{\eta_a}}
; \\
   m\left[2({\eta_a}-\eta_B)  +\cosh^{-1}\left(\frac{m^2}{2 m_a^2}-1\right) \right]  ,
               & \hbox{ if }~
2 m_a c_{{\eta_B}-{\eta_a}} < m 
2 m_a c_{\eta_B}; \\[0.2cm]
   m\left[ 2({\eta_a}+\eta_B) -\cosh^{-1}\left(\frac{m^2}{2 m_a^2}-1\right)\right]  ,
               & \hbox{ if }~
2 m_a c_{{\eta_B}}< m < 2 m_a c_{{\eta_B}+{\eta_a}},
\\
   0, & \hbox{ otherwise },
   \end{array}
 \right. 
\eea
\bea
\left. \frac{d \Gamma}{d m} \right|_{\R_3}
\!\!\!&\propto&
\left\{
   \begin{array}{ll}
   m\left[ 2{\eta_a}-2\eta_B +\cosh^{-1}\left(\frac{m^2}{2 m_a^2}-1\right)\right],
               & \hbox{ if }~
2 m_a c_{{\eta_B}-{\eta_a}} < m <
2 m_a c_{\eta_B}; \\[0.2cm]
   m\left[ 2{\eta_a}+2\eta_B -\cosh^{-1}\left(\frac{m^2}{2 m_a^2}-1\right)\right],
               & \hbox{ if }~ 
2 m_a c_{\eta_B}< m <2 m_a c_{{\eta_B}+{\eta_a}};
\\
   0, & \hbox{ otherwise }.
   \end{array}
 \right.
\eea
Here we have employed the narrow width 
approximation and ignored spin correlation effects.
The detailed derivation is presented in Appendix \ref{sec:invmassderivation}.

\begin{table}[!t]
\centering
\begin{tabular}{cc|cccccc}
\hline
\hline
        &  ~~~Region~~~ & ~~~$m_D$ ~~~ & ~~~$m_B$~~~ &~~~ $m_a$ ~~~ & ~~~$m_X$~~~
            &~~~$\eta_B$~~~ & ~~~${\eta_a}$~~~ \\
\hline
~~~\textsc{Mass--1} ~~~  & $\R_1$ & 650 & 300 & $m_Z$ & 100 & 0.41 & 1.06 \\
~~~\textsc{Mass--2} ~~~   & $\R_2$  & 850  & 330 & $m_Z$ & 100 & 0.74 & 1.18\\
~~~\textsc{Mass--3}  ~~~   & $\R_3$ & 1000 & 250 & $m_Z$ & 100 & 1.32 & 0.80 \\
\hline
\hline
\end{tabular}
\caption{\label{table:antler:mass:Z} Test mass spectrum sets
for the symmetric antler decay.
All masses are in units of GeV and $m_Z$
is the $Z$ boson mass. $\eta_B$ and $\eta_a$
are the rapidities of the particle $B$ and $a$
in its parent rest frame, respectively.}
\end{table}

In order to show the characteristics of the $m$ distribution,
we take three samples for mass parameters in Table \ref{table:antler:mass:Z}.
We label them as  \textsc{Mass--1}, \textsc{Mass--2} 
and \textsc{Mass--3},
each of which corresponds to the kinematical regions of $\R_1$, $\R_2$ and $\R_3$,
respectively.
The visible particle is assumed to be the $Z$ boson.

\begin{figure}[!t]
\centering
{\includegraphics[scale=0.75]{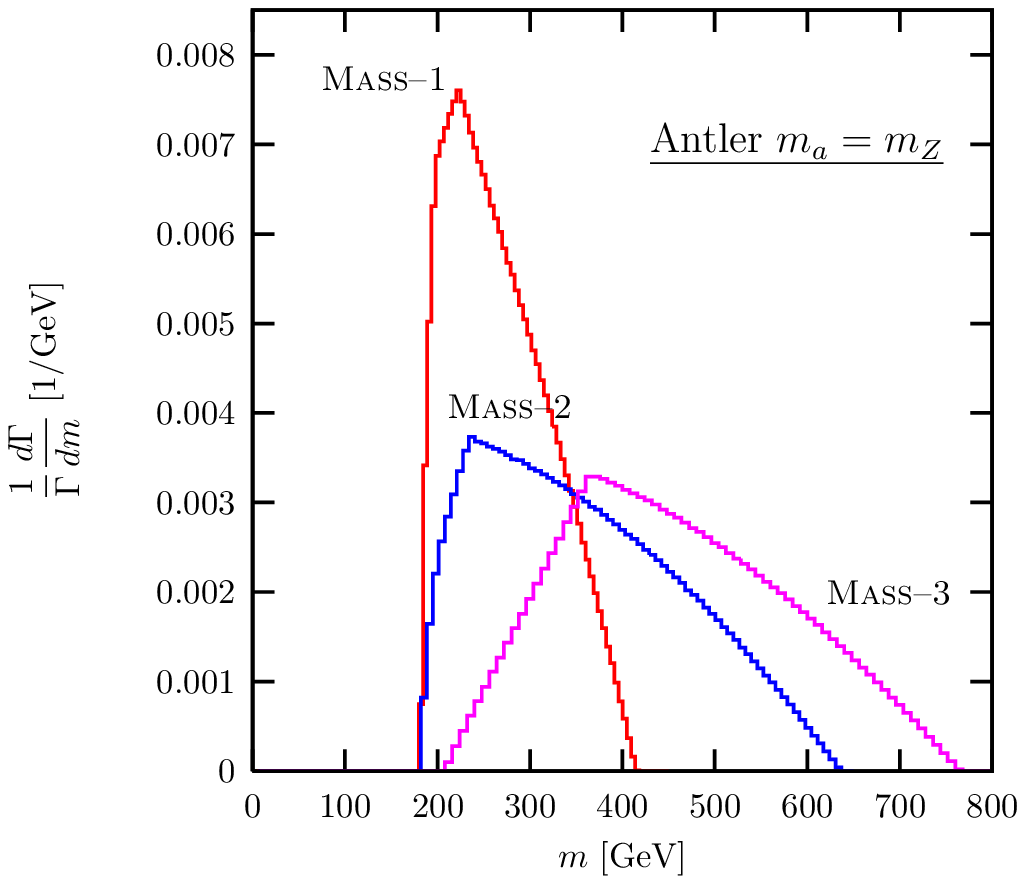}}\
  \caption{The normalized invariant mass distribution $d\Gm/dm$ for test mass sets
in Table \ref{table:antler:mass:Z}.
\label{fig:antler:m} }
\end{figure}

In Fig.~\ref{fig:antler:m},
we present the invariant mass distributions
for the mass parameters in Table \ref{table:antler:mass:Z}.
All three mass sets
yield sharp cusp structures.
The $m$ minimum for $\R_1$ and $\R_2$
is $2 m_Z$ as discussed before.
For the $\R_3$ case,
however, fast-moving intermediate particle $B$
yields $\mmin=2 m_Z \cosh (\eta_B-\eta_a)$.
Unfortunately, we still have a two-fold
ambiguity between $\R_1$ and $\R_2$
because we do not know \textit{a priori} whether the observed
$\mcusp$ is $2 m_a \cosh (\eta_B -\eta_a)$
or $2 m_a \cosh \eta_B$.
As shall be shown in the next section,
the
transverse momentum distribution 
breaks this ambiguity
through its cusp and endpoint structures.
In addition,
the $\R_1$ and $\R_2$ cases have the knee
structure, even though it is challenging to probe
with the expected statistics at the LHC. 

\subsection{Transverse momentum variables: $m_T$, $p_T$, and $p_{Ti}$}
\label{subset:massive:pt}

\begin{figure}[!t]
\centering
  \includegraphics[scale=0.75]{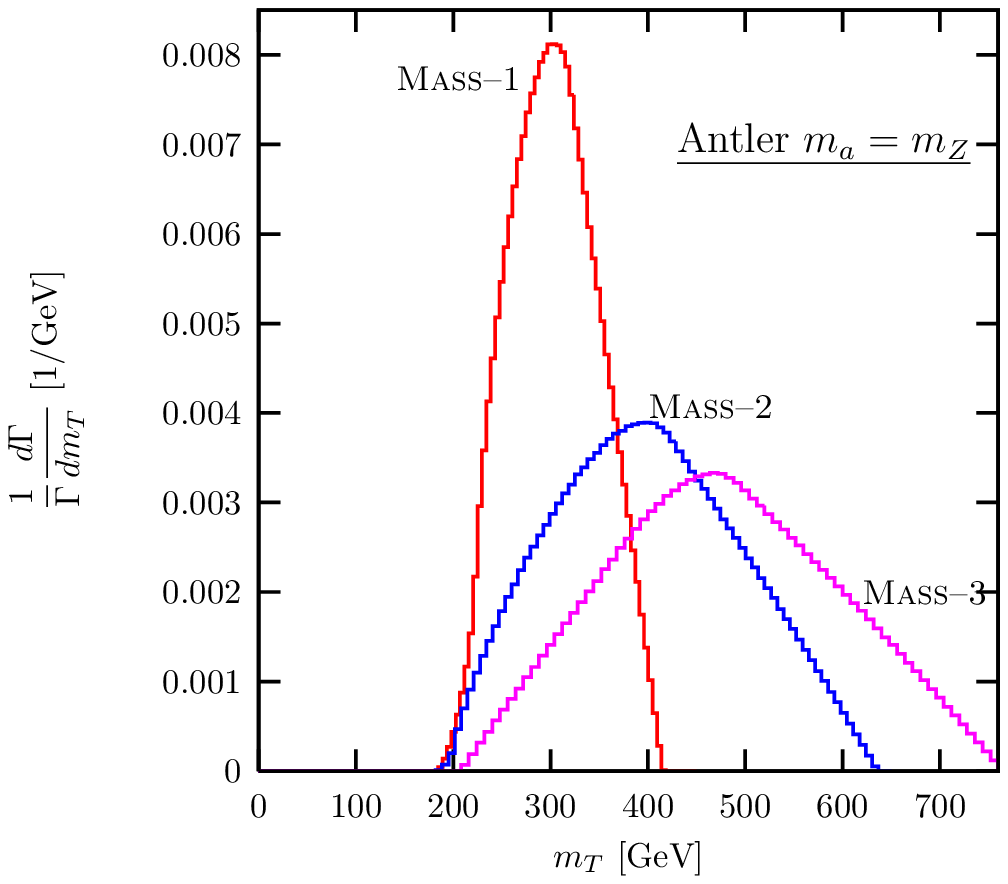}
  \caption{\label{fig:antler:mT}
  The normalized  transverse mass distribution $d\Gm/dm_T$ for test mass sets
in Table \ref{table:antler:mass:Z}.}
\end{figure}

In this section, we investigate the distribution 
of the transverse mass $m_T$,
 the transverse momentum variables $p_{T}$ and 
$p_{Ti}$.
In Fig.~\ref{fig:antler:mT}, we show
the $m_T$ distribution.
All the $m_T$ distributions for $\R_1$, $\R_2$, and $\R_3$
do not have any cusped peak.
The maximum in the $m_T$ distribution
is the same as the maximum of $m$:
\bea
\label{eq:mt:max}
\left( m_T \right)_{\rm max} = m_{\rm max}.
\eea
The confirmation of the same maxima
in the $m$ and $m_T$ distributions
will help the reconstruction of the antler decay. 

\begin{figure}[!t]
\centering
{\includegraphics[scale=0.75]{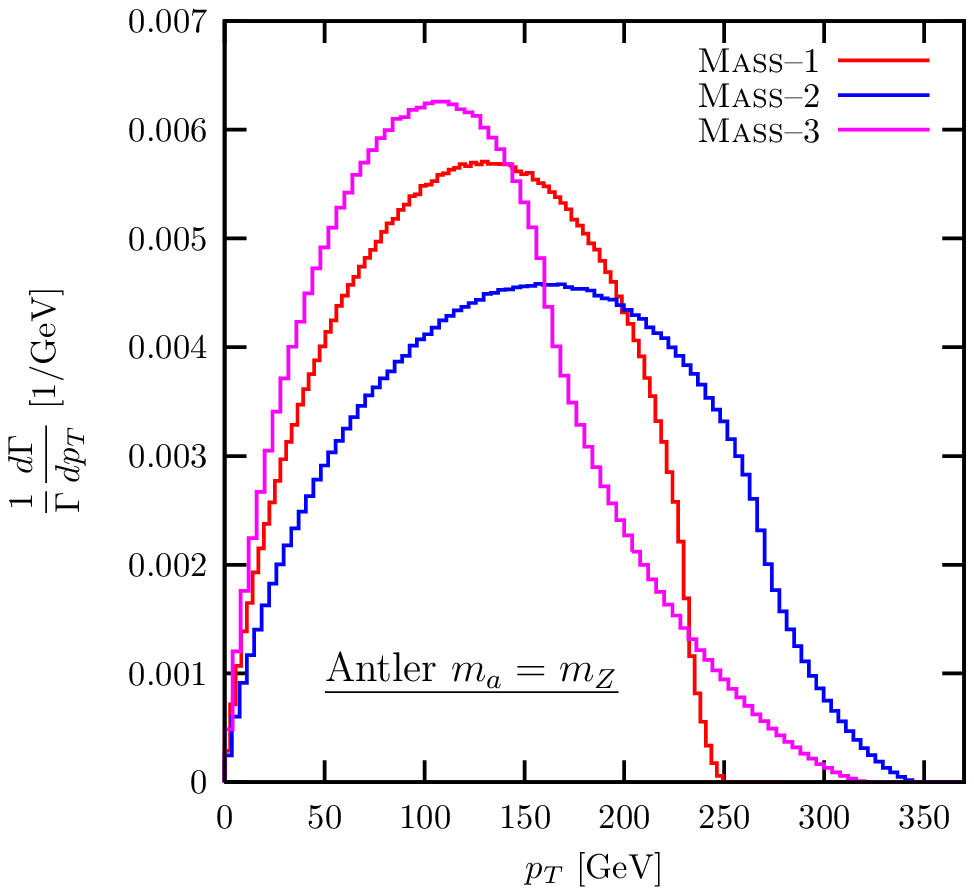}}\quad
{\includegraphics[scale=0.75]{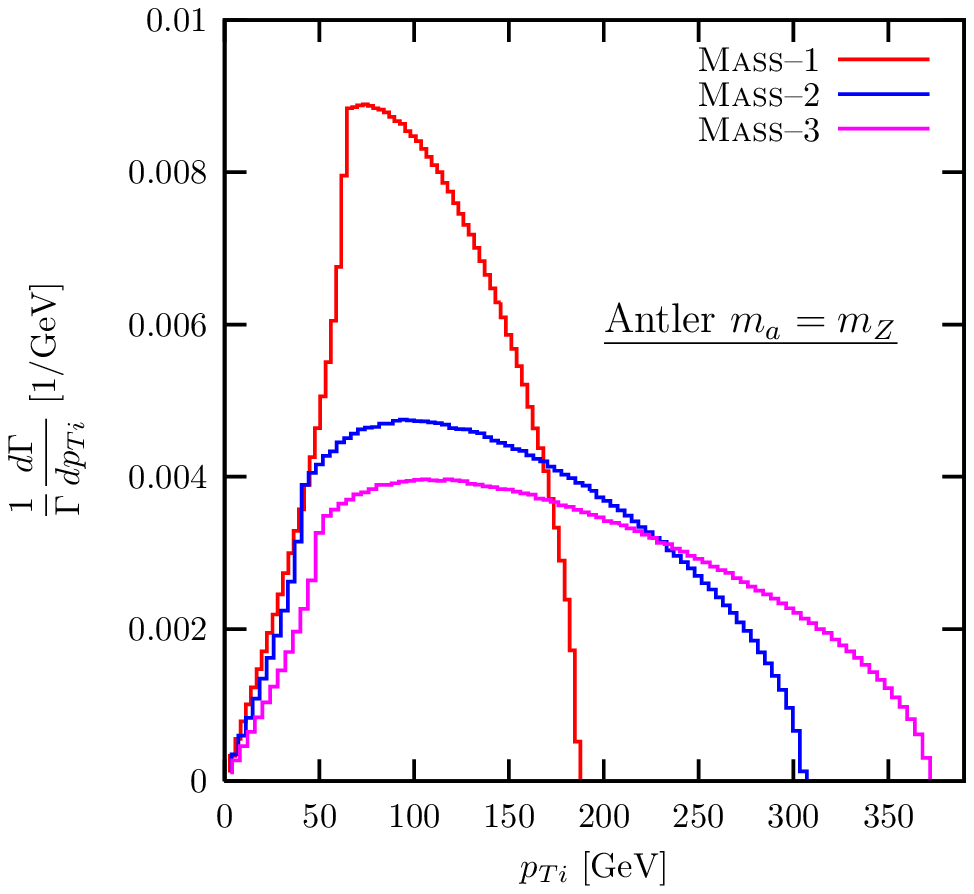}}
  \caption[The differential decay width $d\Gm/d p_T$ 
and $d \Gamma /d p_{Ti}$ for the mass spectrum]
{\label{fig:antler:pt:pti}
The normalized transverse momentum distribution $d\Gm/d p_T$ and $d \Gamma /d p_{Ti}$ 
for test mass sets in Table \ref{table:antler:mass:Z}.
}
\end{figure}

In Fig.~\ref{fig:antler:pt:pti}, we plot the distribution
of $p_T$
and $p_{Ti}$.
The total $p_T$ distribution does not 
reveal the cusp structure,
as expected from the $m_T$ distribution.
In addition, its maximum is at the end of a long tail,
which is statistically disadvantageous to observe.
The cluster transfer mass $m_C$ of $a_1$-$a_2$-$\,  \ptmiss$ system
has no cusp structure either.

The transverse momentum of ``one" visible particle shows quite different 
distribution.
First, note that one unambiguous $p_{Ti}$ distribution
can be constructed out of two visible particles,
because of the symmetric topology of the antler decay.
This $p_{Ti}$ distribution
shows the cusp structure 
as well as the fast-dropping maximum structure.
The cusp and maximum of $p_{Ti}$ are
\bea
\label{eq:pti:massive}
\left(p_{Ti}\right)_{\rm cusp} &=& m_a \left| \sinh ({\eta_a}-\eta_B) \right|,
\\ \no
\left(p_{Ti}\right)_{\rm max} &=& m_a \sinh \left( \eta_a+{\eta_B} \right).
\eea
Note that $\left(p_{Ti}\right)_{\rm max}$ gives the information
about $\eta_B +\eta_a$, which is the same from
$m_{\rm max}$ in Eq.~(\ref{eq:mmax}).
Remarkable is that
$\left(p_{Ti}\right)_{\rm cusp}$ is common for all three regions $\R_{1,2,3}$,
which determines $|\eta_B -\eta_a|$.
By comparing $(p_{Ti})_{\rm cusp}$
with $m_{\rm cusp}$,
we can distinguish $\R_1$ from $\R_2$.
This breaks the two-fold ambiguity in the
measurement of $m_{\rm cusp}$ for $\R_1$ and $\R_2$.

\subsection{Angular variable: $\cos \Theta$  }
\label{subsec:massive:cos}

\begin{figure}
\centering
{\includegraphics[scale=0.75]{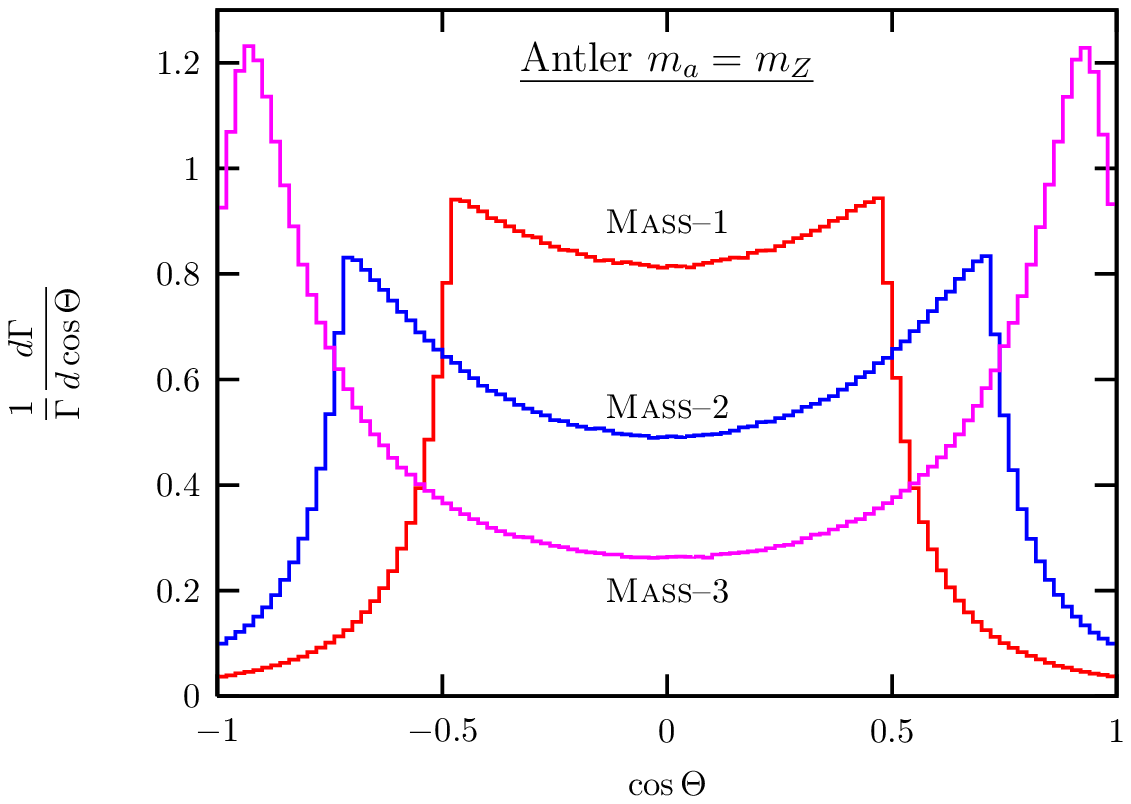}}
  \caption[The $\cos\Th$ distribution
  for the massive visible particle cases ]{\label{fig:antler:costh}
The normalized $\cos\Th$ distribution
  for the massive visible particle cases.
 }
\end{figure}

We consider the distribution of $\cos\Theta$ defined in Eq.~(\ref{eq:def:costh}).
Here $\Theta$ is the angle of one visible particle 
with respect to the c.m. moving direction.
As in the $p_{Ti}$ distribution,
the symmetric decay chains of the antler decay
guarantee one unique $\cos\Th$ distribution
as shown in Fig.~\ref{fig:antler:costh}.
All $\cos\Theta$ distributions for $\R_1$, $\R_2$ and $\R_3$
are symmetric about $\cos\Theta=0$, and
have sharp cusps.

\section{Massless visible particle case}
\label{sec:massless}

Now we consider the massless visible particle case.
As suggested in Eqs.~(\ref{eq:zp:susy}), (\ref{eq:mued}), and (\ref{eq:ilc}),
many new physics processes for the antler decay have massless visible particles.
Although we cannot 
directly apply the results 
with the massive visible particle
to this case since the rapidity $\eta_{a}$
diverges, 
we can obtain the massless limit by using the finite combinations of 
$m_a c_{\eta_a}$ and $m_a s_{\eta_a}$:
\bea
\label{eq:massless:limit}
\lim_{m_a\to 0}
m_a
\chz
&=&
\lim_{m_a\to 0}
m_a \shz
=\frac{m_B}{2}
\left(1-\frac{m_X^2}{m_B^2}
\right).
\eea

\subsection{Invariant mass distribution}
\label{subsec:massless:m}
In the massive visible particle case, 
the functional form of the invariant mass distribution is different
according to three mass regions of ${\R_1}$, ${\R_2}$, and ${\R_3}$.
In the massless visible particle case, 
only $\R_1$ applies since $\eta_B \ll \eta_a$. 
Two locations of $\mmin$ and $\mknee$ merge because $m_a=0$.
The cusp and endpoints are given by 
\begin{eqnarray}
m_{\rm min}^{(0)} &=& 0, \\ 
m_{\rm cusp}^{(0)} &=& m_B \left(1-\frac{m_X^2}{m_B^2}\right)
e^{- \eta_B} 
, \label{eq:masscusp0}
\\
m_{\rm max}^{(0)} &=&  m_B \left(1-\frac{m_X^2}{m_B^2}\right)
e^{\eta_B}
.
\label{eq:massmax0}
\end{eqnarray}
Here the superscript $(0)$ is used for emphasizing 
$m_a=0$. 
The product 
of the cusp and the maximum is
\bea
m_{\rm cusp}^{(0)}  m_{\rm max}^{(0)} 
= {m_B^2}\left(1-\frac{{m_X^2}}{m_B^2}\right)^2,
\eea
which depends only on the second step decay of $B \to a X$.
The ratio is
\bea
\frac{m_{\rm cusp}^{(0)}}{m_{\rm max}^{(0)}}
= e^{-2\eta_B},
\eea
which is determined only by the first step decay
of $D \to BB$.

The invariant mass distribution is simplified into
\begin{eqnarray}
\label{eq:dGmdm:0}
\frac{d\Gamma}{dm} 
\propto
\begin{cases}
\begin{array}{ll}
m \log \left( \frac{m_{\rm max}^{(0)}}{m_{\rm cusp}^{(0)}} \right), &  
\mbox{ if $0 < m < m_{\rm cusp}^{(0)}$; } \\
m \log \left( 
\frac{m_{\rm max}^{(0)}}{m}
\right),
& 
\mbox{ if $ m_{\rm cusp}^{(0)} < m  < m_{\rm max}^{(0)}$;} \\
0, & \mbox{ otherwise. } 
\end{array}
\end{cases}
\end{eqnarray}
For $0 < m < m_{\rm cusp}^{(0)}$,
$ d \Gamma/dm$ is a linear function of $m$.
For $m_{\rm cusp}^{(0)} < m < m_{\rm max}^{(0)}$,
it is a concave function with the maximum at 
$m=m_{\rm max}^{(0)}/e$.
Depending on the relative position of $m_{\rm cusp}^{(0)}$
and $m_{\rm max}^{(0)}/e$,
the maximum of the concave function may or may not show
in the function of $d \Gm/d m$,
which determines 
the sharpness of the cusp.
If $m_{\rm max}^{(0)}/e < m_{\rm cusp}^{(0)}$ 
(or equivalently $m_B > 0.443\, m_D$),
$d \Gm/d m$ is linearly increasing up to $m=m_{\rm cusp}^{(0)}$,
and  decreasing after that:
the cusp is sharp.
If $m_{\rm cusp}^{(0)}< m_{\rm max}^{(0)}/e$,
$d \Gm/d m$ keeps increasing after $m=m_{\rm cusp}^{(0)}$,
reaches the maximum of the concave function,
and finally falls down:
the cusp is not sharp.
The $m$ cusp structure is most useful
when the $D \to BB$ decay is near the threshold.

\begin{table}
\centering
\begin{tabular}{cccccccc}
\hline
\hline
         & ~~$m_D$~~ & ~~$m_B$~~ & ~~$m_X$~~&~~$m_a$~~
            & ~~$m_{\rm cusp}^{(0)}$~~  & ~~$m_{\rm max}^{(0)}$~~
            &  ~~$|\cos\Th|_{\rm max}$~~ \\
\hline
~~~\textsc{Mass--1}$_0$~~~  & 1000 & 470 & 440 & 0 & 40.7 &82.9 &  0.34 \\
~~~\textsc{Mass--2}$_0$~~~  & 1000 & 440 & 410 & 0 & 34.6 & 97.1 & 0.47\\
~~~\textsc{Mass--3}$_0$~~~  & 1000 & 400 & 370 & 0 & 28.9 & 115.5 &  0.60 \\
\hline
\hline
\end{tabular}
\caption{\label{table:antler0:mass} Test mass spectrum sets
for the symmetric antler decay with massless SM particles.
All masses are in units of GeV.
}
\end{table}

In order to show the functional behaviors specifically,
we take three mass sets for the massless visible particle case 
in Table \ref{table:antler0:mass}.
The mass parameters in the \textsc{Mass--1}$_0$
correspond to the case where both the first decay $D \to BB$
and the second decay $B \to aX$ occur near the threshold.
This is motivated 
by the decay of the second KK mode of $Z$ boson 
in the UED model in Eq.(\ref{eq:mued}).
The \textsc{Mass--2}$_0$ represents
the marginal case for the sharp cusp, \textit{i.e.},
$m_B \approx 0.44\ m_D$.
The \textsc{Mass--3}$_0$ case has large mass gaps.

\begin{figure}[!t]
\centering
{\includegraphics[scale=0.75]{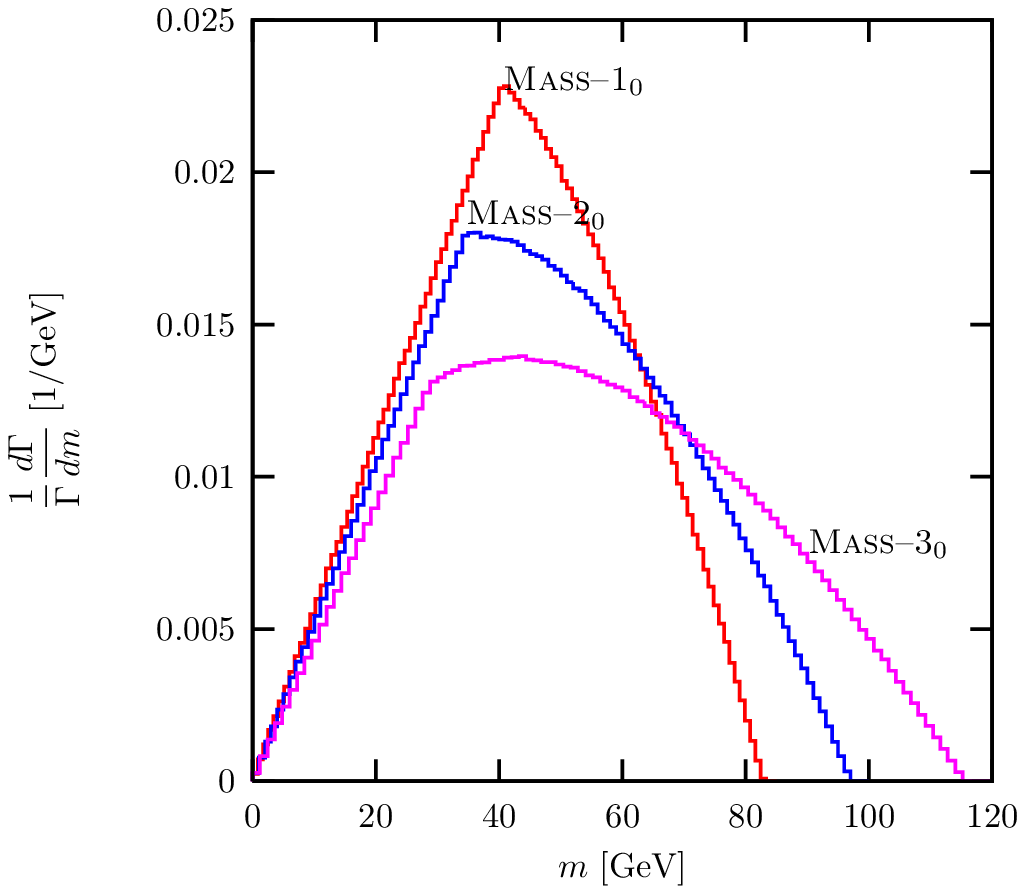}}
  \caption[The invariant mass distribution $d\Gm/dm$ 
for test mass spectrum sets]
{The normalized invariant mass distribution $d\Gm/dm$ in the massless visible 
particle case for the mass parameter sets in Table \ref{table:antler0:mass}. \label{fig:antler0:m} }
\end{figure}

Figure \ref{fig:antler0:m} shows the $m$ distributions.
All three mass sets in Table \ref{table:antler0:mass}
have the same $\mmin=0$.
The sharpness of the cusp structure is
different.
The nearly degenerate mass case (\textsc{Mass--1$_0$})
has a very sharp cusp.
The marginal case  (\textsc{Mass-2$_0$})
shows also an observably sharp cusp.
The large mass gap case (\textsc{Mass-3$_0$})
has a rather smooth cusp.
If the number of events is not enough,
the obtuse cusp in the  \textsc{Mass-3$_0$}
is difficult to read.
The measurement of the $m$ cusp 
is still possible
since the functional form in Eq.~(\ref{eq:dGmdm:0})
can be used to fit the data and to read the cusp position.

\subsection{Transverse momentum variables: $m_T$, $p_{T}$ and $p_{Ti}$} 
\label{subsec:massless:pt}

\begin{figure}[!t]
\centering
  \includegraphics[scale=0.75]{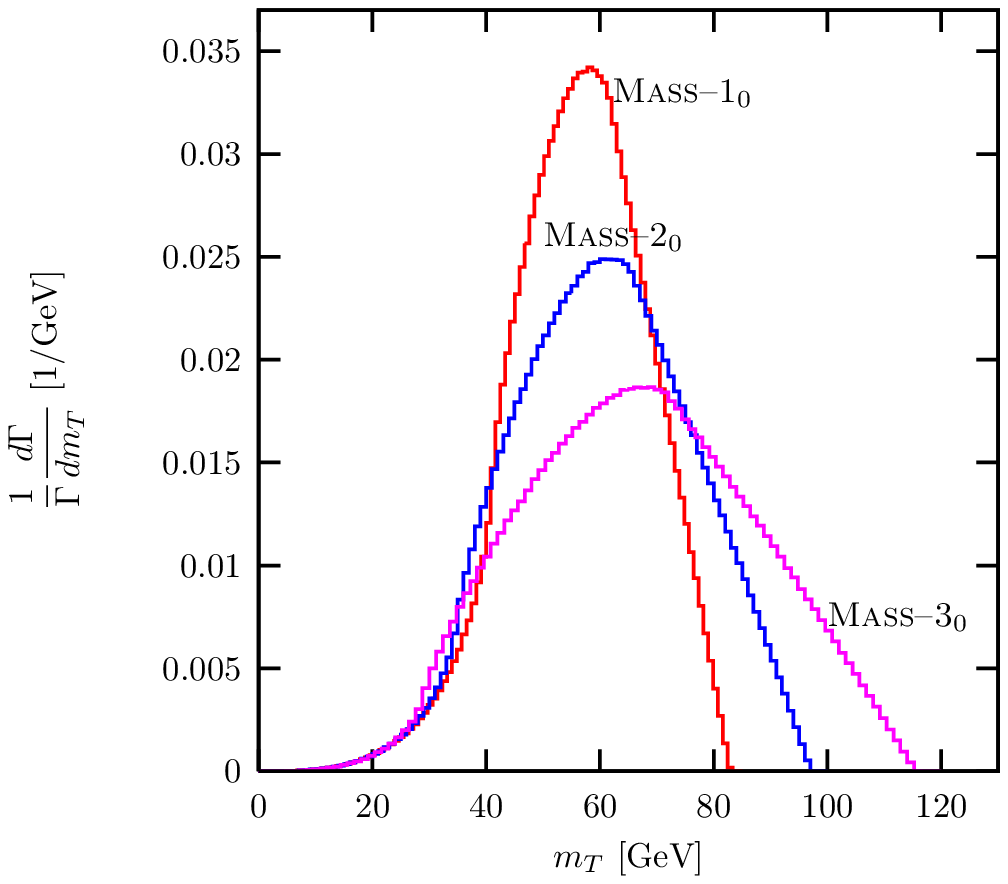}
  \caption[1]{\label{fig:antler0:mT}
The normalized transverse mass distribution 
$d\Gm/d m_T$ for the massless visible particles.
The mass spectrum sets are described in
Table \ref{table:antler0:mass}.}
\end{figure}

Now we turn to the kinematic variables involving
transverse monentum.
First, 
the $m_T$ distribution in the massless visible particle case
does not show any cusp structure
as shown
in Fig.~\ref{fig:antler0:mT}.
The absence of $m_T$ cusp is 
a common feature of the antler decay.
The $m_T$ maximum stands at the end of fast-dropping function
for all three mass sets,
which is easier to read.
In addition it
is the same as the $m$ maximum:
\bea
(m_T)_{\rm max}^{(0)} = \mmax^{(0)}.
\eea

\begin{figure}
\centering
{\includegraphics[scale=0.75]{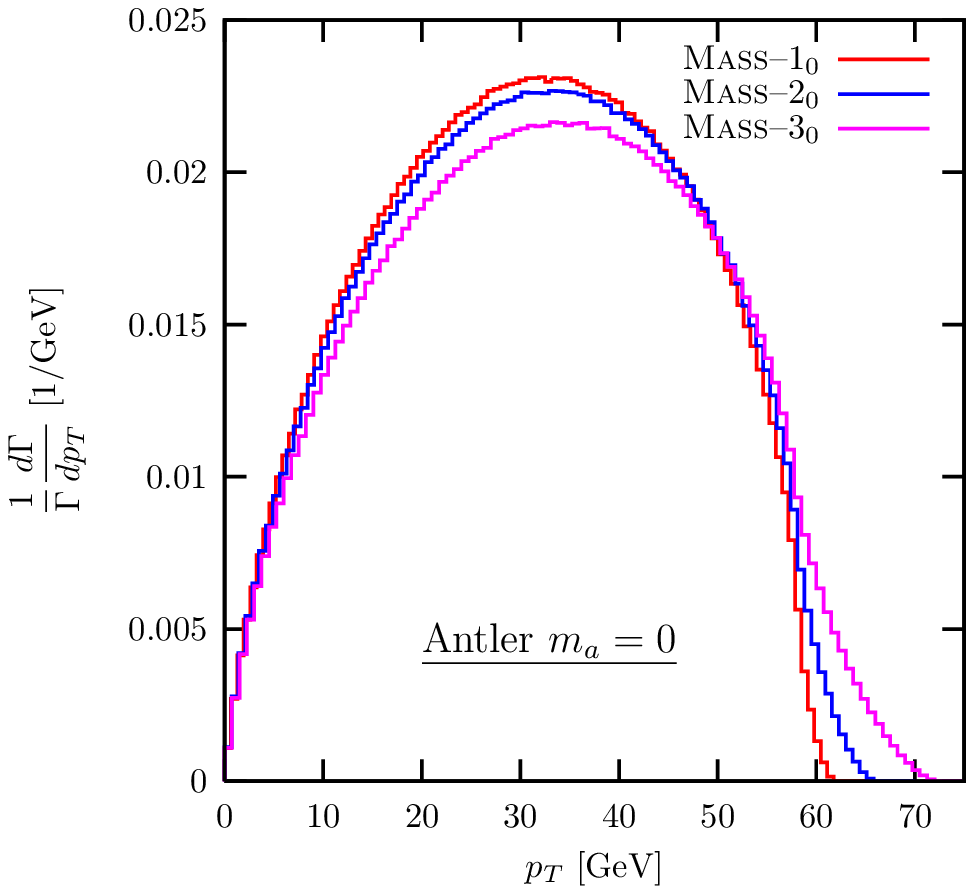}}\quad
{\includegraphics[scale=0.75]{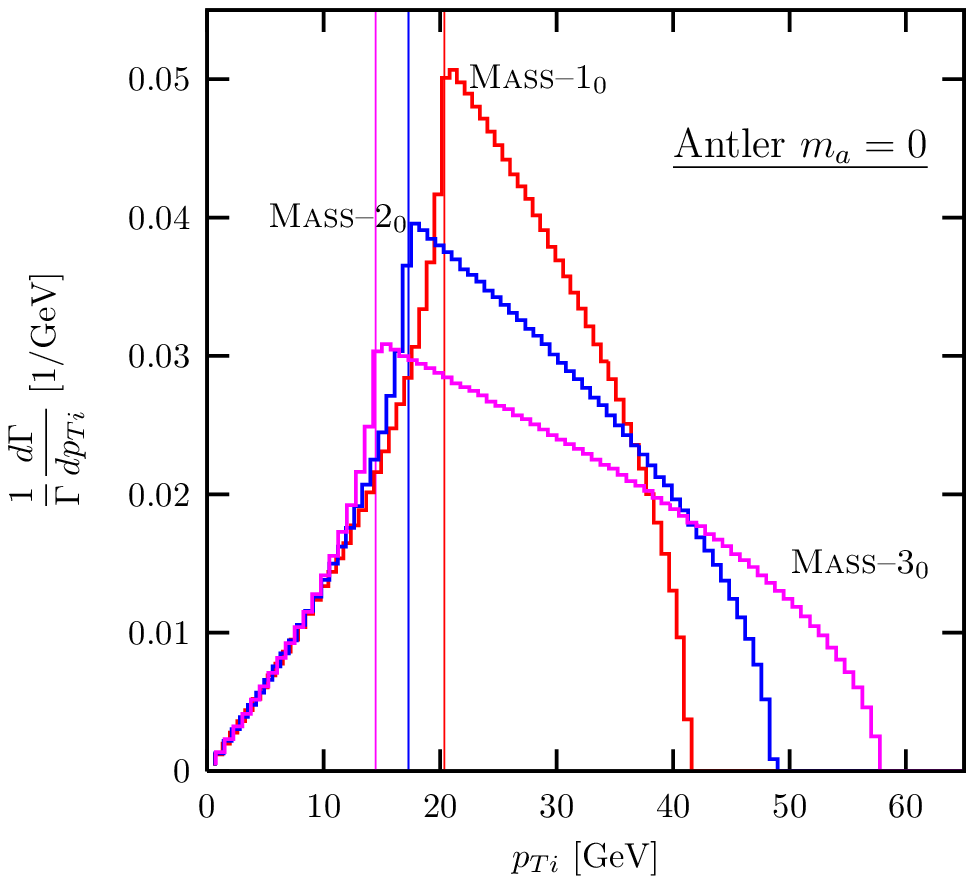}}
  \caption[The transverse momentum distribution $d\Gm/d p_T$ 
for the massless visible particles.]{\label{fig:antler0:pT;pTi}
The normalized transverse momentum distribution $d\Gm/d p_T$ and  $d\Gm/d p_{T_i}$  
for the massless visible particles.
Test mass sets are in
Table \ref{table:antler0:mass}.}
\end{figure}

Figure \ref{fig:antler0:pT;pTi} shows the distributions of
the total transverse momentum $p_T$ and individual $p_{Ti}$.
As in the massive visible particle case,
the total $p_T$ distribution is very smooth and gentle, 
without any cusp structure
or fast dropping maximum.
Instead, the $p_{Ti}$ distribution shows very sharp cusp,
much sharper in general than the 
invariant mass distribution.
Even the \textsc{Mass--3}$_0$ case,
which suffers from the dull cusp 
in the $m$ distribution,
has a very sharp $p_{Ti}$ cusp.
In addition the $p_{Ti}$ maximum 
is at the end of a faster
dropping function.

The analytic expressions 
of $\left( p_{T_i} \right)^{(0)}_{\rm cusp}$ and 
$\left( p_{T_i} \right)^{(0)}_{\rm max}$
can be easily obtained from Eq.~(\ref{eq:pti:massive})
by applying Eq.~(\ref{eq:massless:limit}):
\bea
\label{eq:pti:cusp:massless}
\left( p_{Ti} \right)_{\rm cusp}^{(0)} = 
\frac{1}{2} m_{\rm cusp}^{(0)}, \quad
\left( p_{Ti} \right)_{\rm max}^{(0)} = 
\frac{1}{2} m_{\rm max}^{(0)}. 
\eea
The measurements of 
$\left(p_{Ti}\right)_{\rm cusp}^{(0)}$ and 
$\left(p_{Ti}\right)_{\rm max}^{(0)}$ provide
the same information as $m_{\rm cusp}^{(0)}$ and 
$m_{\rm max}^{(0)}$,
which is another way to check the antler decay topology.

\subsection{Angular variable: $\cos \Theta$  }
\label{subsec:massless:cos}

\begin{figure}
\centering
{\includegraphics[scale=0.75]{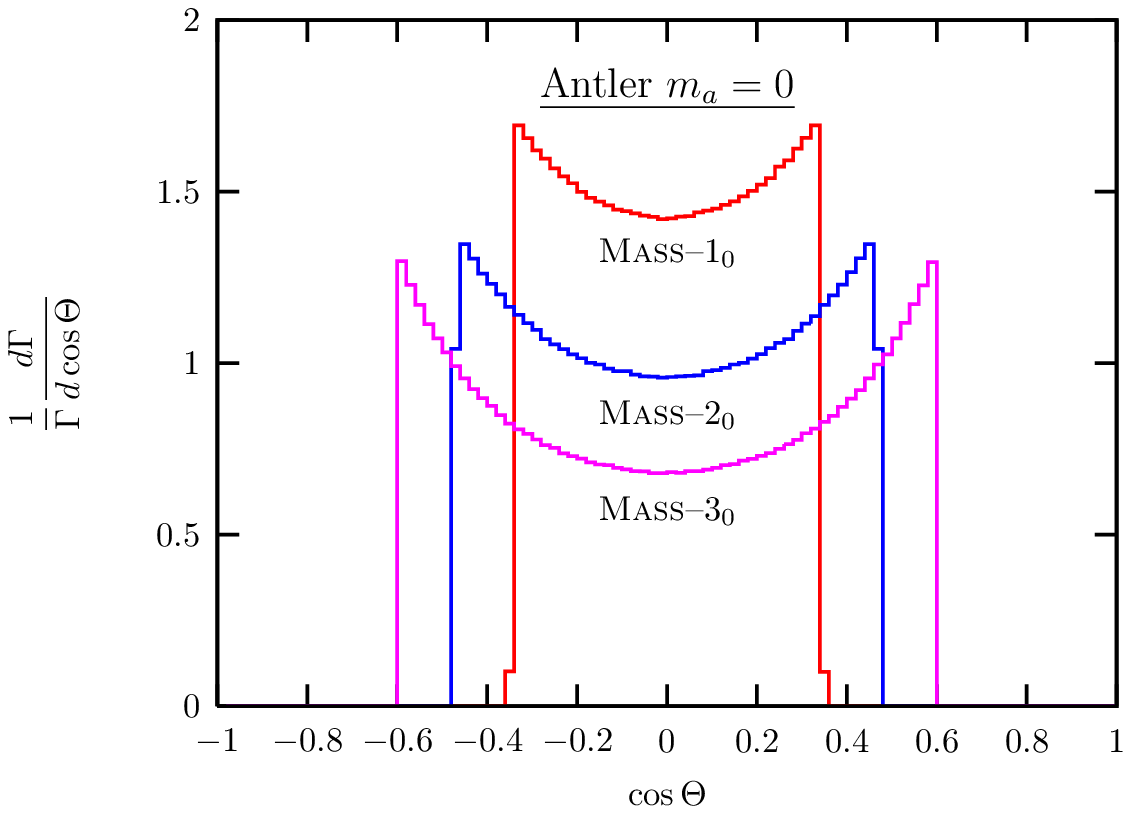}}  
 \caption[The $\cos\Th$ distribution: massless visible particle cases (a) and massive visible particle cases (b) ]{\label{fig:antler0:costh}
The normalized $\cos\Th$ distribution
for the massless visible particle cases. 
 }
\end{figure}

Figure \ref{fig:antler0:costh}
shows the  normalized
$d \Gm/ d \cos\Th$ distributions
for three massless visible particle cases.
The function 
increases with $|\cos\Th|$,
and drops to zero suddenly at 
$\left| \cos\Th  \right|^{(0)}_{\rm max}$.
This is because the cusp and the endpoint merge, 
resulting in more pronounced endpoints with sharp peaks at both ends. 
The maximum of $\cos\Th$ is simply
determined by the first step decay $D \to B B$:
\bea
\label{eq:max:costh0}
\left| \cos\Th  \right|^{(0)}_{\rm max} = \tanh\eta_B.
\eea
The full analytic function of $\dphif /d \cos\Th$ is given by
\bea
\label{eq:costh0:distribution}
\frac{\Gm}{d \cos\Th}
\propto 
\left\{
\begin{array}{ll}
\dfrac{1}{\sin^3\Th}, & \hbox{ for } |\cos\Theta|< \tanh \eta_B,
\\
0, & \hbox{ otherwise. } 
\end{array}
\right.
\eea
The suddenly ending behavior of the $\cos\Theta$ distribution is
because massless visible particles cannot access all kinematic space of $\cos\Theta$.
The detailed derivation
of Eqs.~(\ref{eq:max:costh0}) and
(\ref{eq:costh0:distribution})
is in Appendix \ref{appendix:cosTh}.

\section{Effects from realistic considerations} 
\label{sec:real:effects}

In the previous sections, 
we have considered the kinematics only, 
ignoring the decay width of 
the intermediate particle $B$, the longitudinal boost 
of the parent particle $D$, and 
the spin correlation. 
These $S$-matrix element effects
 can smear the kinematic cusps and endpoints.
In the following, we discuss the limitation of 
determining the missing particle mass
using kinematic singularities.

\subsection{Finite width effects}
\label{subsec:width}

The previous results are based on the narrow width approximation.
This approach is very effective
for the proposed processes in the MSSM, 
$Z'$ supersymmetry, UED, and LHT models
since
all the intermediate particles 
($\neu_2$, $\tilde{\ell}^\pm$, $L^\on$, and $t_-$)
have very small total decay widths, 
much smaller than one percent of their masses.
If the total decay width $\Gm_B$ is large,
its effects can smear the cusp and endpoint structures.
If  
the on-shell $B$ particle is kinematically not accessible so that
the decay process is through off-shell $B$, then 
the singular structures are destroyed completely since there
is no constraints on the phase space from the mass relations.

\begin{figure}
\centering
{\includegraphics[scale=0.75]{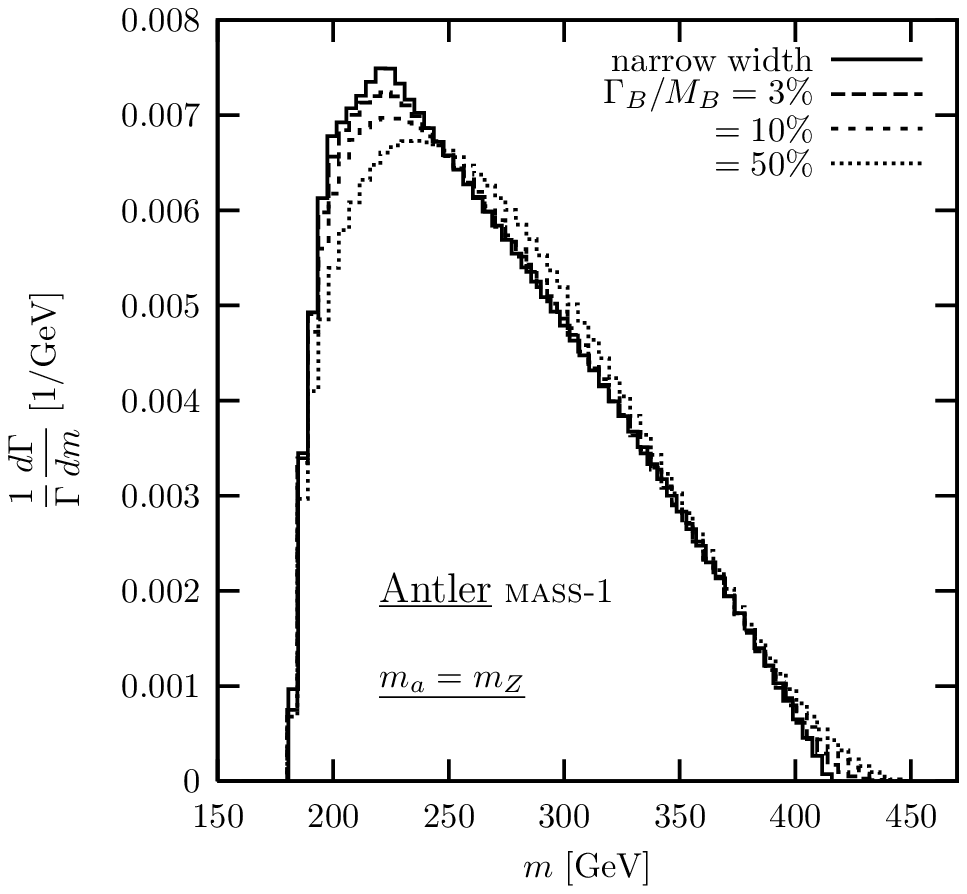}}
{ \includegraphics[scale=0.75]{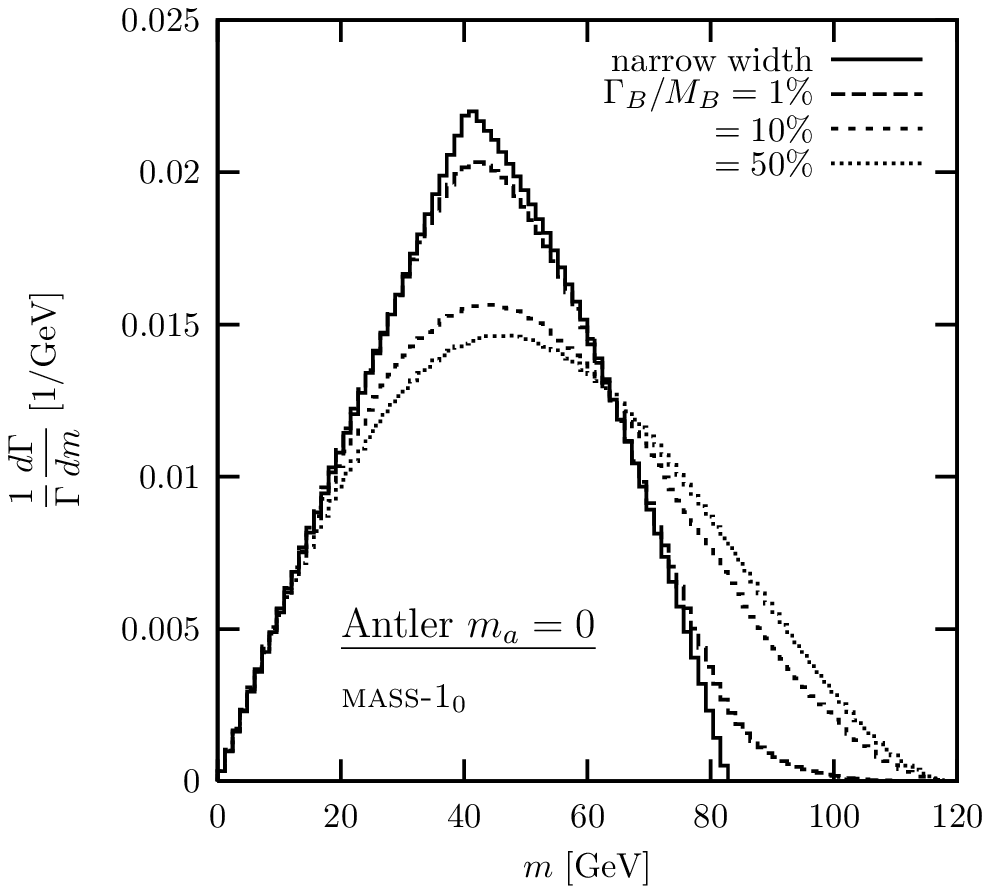}}
  \caption[The invariant mass distribution $d\Gm/d m$ 
in the antler decay with various finite total decay width]
{\label{fig:ant:m:width}
The invariant mass distribution $d\Gm/d m$ 
in the antler decay with finite $\Gm_B$ effects.
We have taken the mass spectrum sets 
of \textsc{Mass--1} and \textsc{Mass--1}$_0$.}
\end{figure}

In Fig.~\ref{fig:ant:m:width},
we show the invariant mass distributions
with the effect of finite $\Gm_B$ 
for the massive SM particle case (\textsc{Mass--1})
and the massless case (\textsc{Mass--1}$_0$).
We take $\Gm_B$ to be 3\%, 10\%, and  50\%
of $m_B$ for the massive case,
and 1\%, 10\%, and 50\% for the massless case.
If $\Gm_B/M_B$ is small enough ($\leq 3\%$ for the massive case
and $\leq 1\%$ for the massless case),
the $m$ cusp remains fairly preserved. 
Even though the sharp cusp gets dull slightly,
the position of the cusp is not shifted for both cases.
The endpoint position is stable
for the massive case, but 
shifted considerably for the massless case.
If $\Gm_B/m_B$ is about 10\%,
the cusp is smeared into a round peak
and the endpoint position 
is shifted significantly
for both cases.
Still the peak of the smeared cusp
stands at the same cusp position.
If $\Gm_B /M_B\simeq 50\% $ in which case a large contribution to
the $S$-matrix element arises from the intermediate off-shell $B$,
the sharpness and position of the cusp are lost. 
The endpoints move towards new positions of
$m = m_D - 2 m_X$.
This is from the allowed
phase space of the decay $D \rightarrow X X a a$.
At least we can determine the mass difference
between $D$ and $X$ using the $m$ distribution.

\begin{figure}[!t]
\centering
{\includegraphics[scale=0.75]{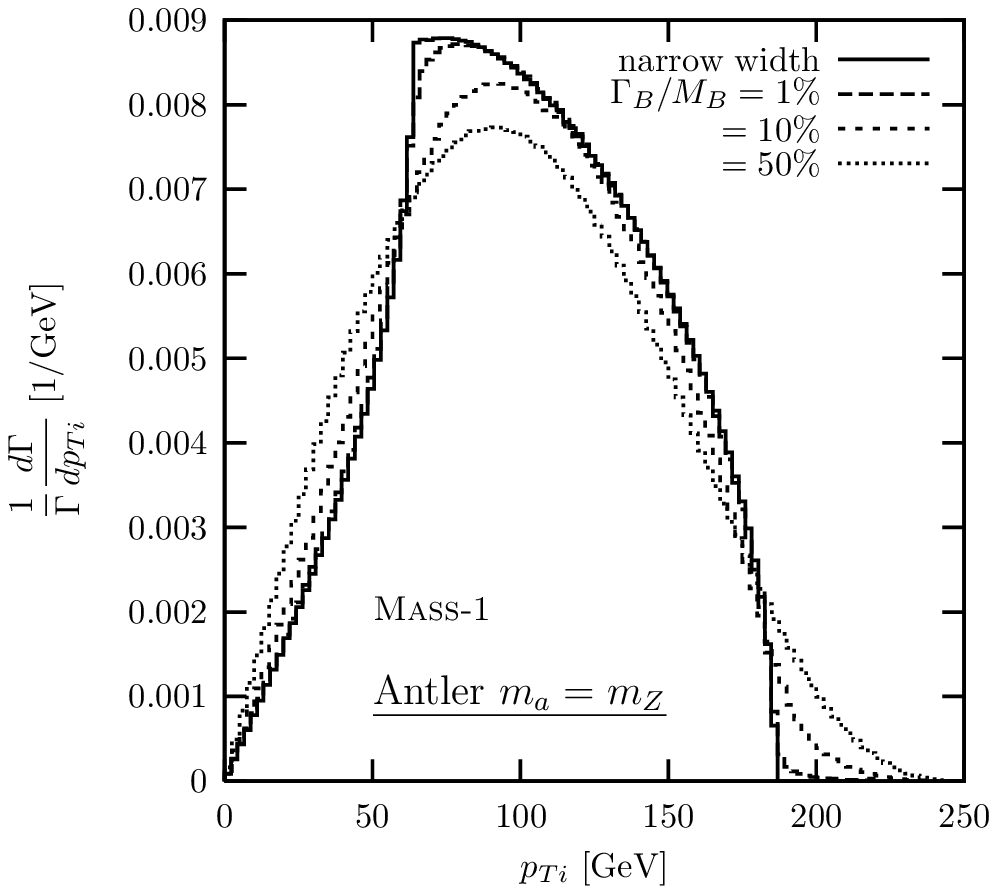}}
{ \includegraphics[scale=0.75]{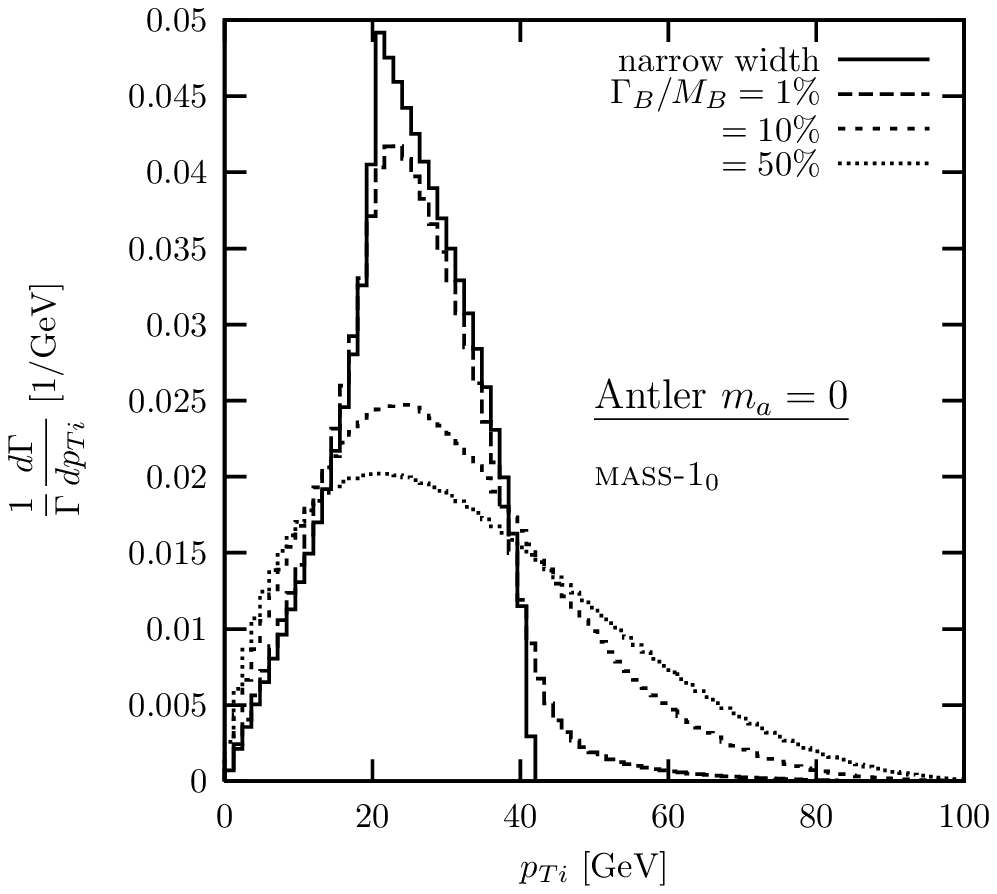}}
  \caption[The transverse momentum distribution $d\Gm/d p_T$]{\label{fig:ant:pt:width}
The normalized transverse momentum distribution 
$d\Gm/d p_T$ 
with various finite total decay width
of the intermediate particle $B$.}
\end{figure}

Now we show the $\Gm_B$ effects
on the $p_{Ti}$ distributions
in Fig.~\ref{fig:ant:pt:width}.
We take 
the massive \textsc{Mass--1} case and
the massless \textsc{Mass--1}$_0$ case.
The $p_{Ti}$ distribution has very vulnerable cusp and endpoint
from the finite $\Gm_B$ effects.
Even for small width effects 
($\leq 3\%$ for the massive case
and $\leq 1\%$ for the massless case)
the sharp cusp becomes dull, and
its position is significantly shifted.
The $p_{Ti}$ maximum is more sensitive
to the $\Gm_B$ effects.
For the massless SM particle case,
even $1\%$ of $\Gm_B/m_B$ 
shifts the position of $ p_{Ti}^{\rm max}$ a lot.

\begin{figure}
\centering
\includegraphics[scale=0.75]{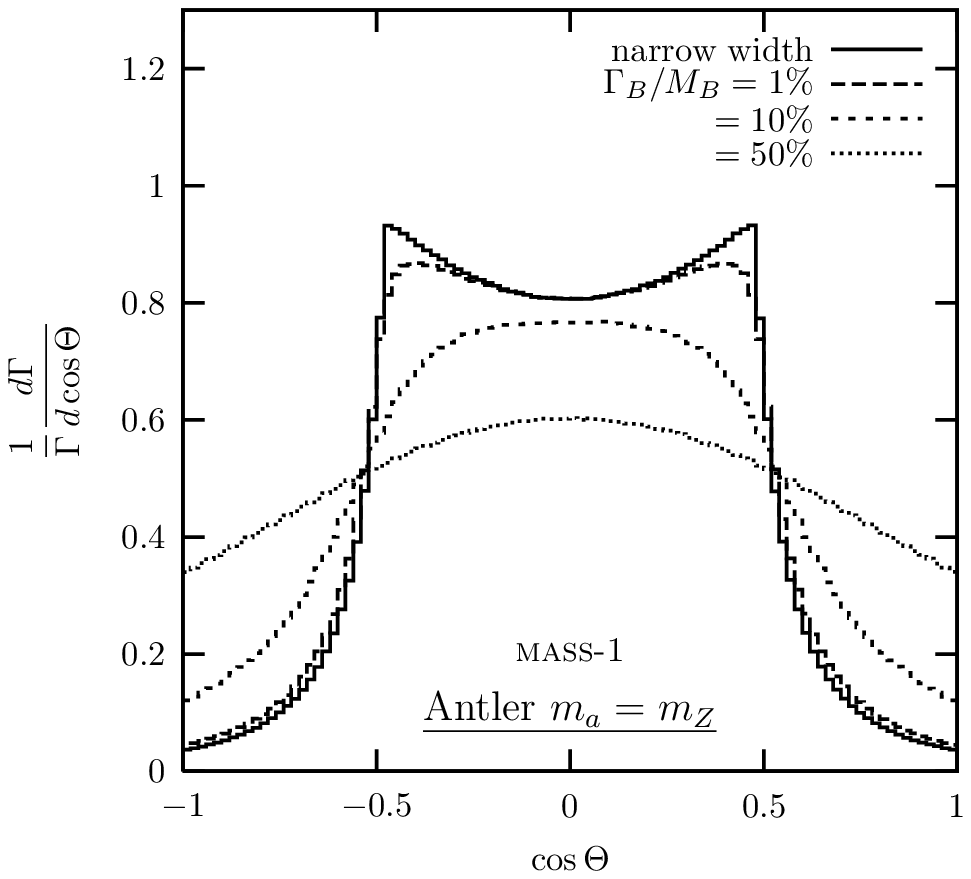}
\includegraphics[scale=0.75]{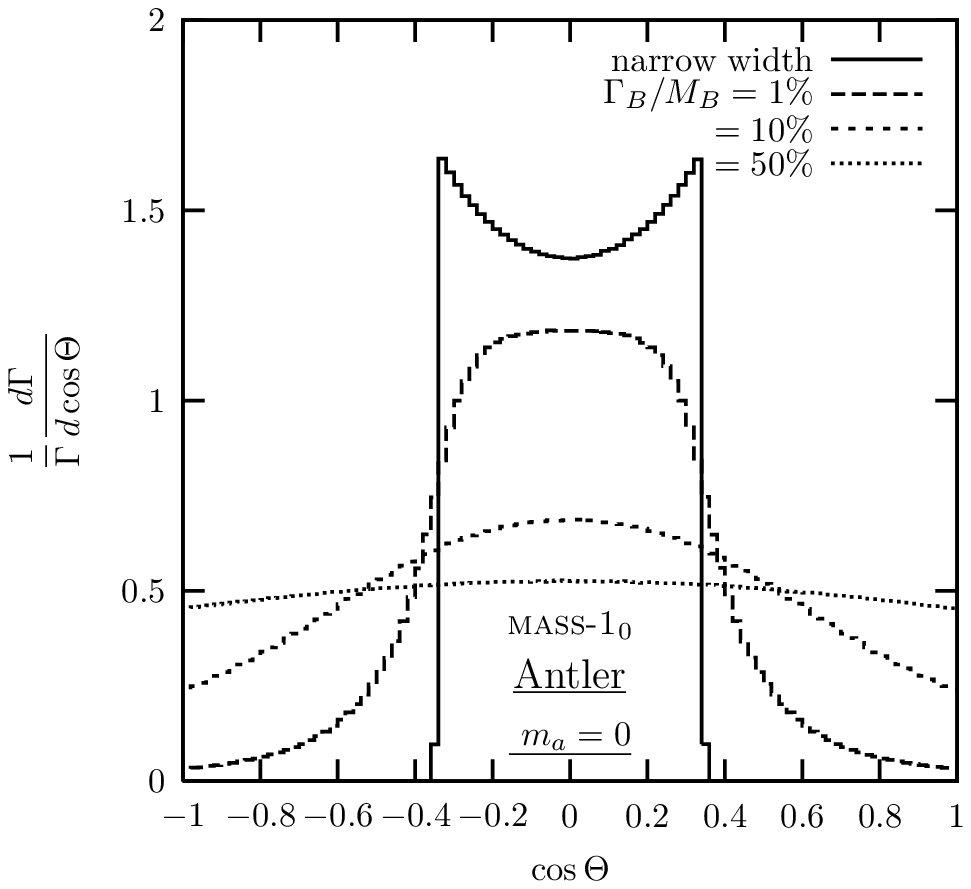}
\caption{\label{fig:ant:costh:width}
The normalized $\cos\Th$ distribution 
with various finite total decay width
of the intermediate particle $B$.}
\end{figure}

Finally, Fig.~\ref{fig:ant:costh:width}
shows the $\cos\Th$ distribution
with finite $\Gm_B$ effects.
Here the most dramatic collapse occurs.
Even with very small width of $\Gm_B/m_B =1\%$,
the sharp cusp 
becomes round,
difficult to read.
For $\Gm_B/ m_B =10\%$,
the cusp shape is lost completely.

In summary, the effects of the
finite width of the intermediate particle $B$ 
smear the cusp shape and shift the cusp position to some extent.
The invariant mass distribution has the least 
distortion, while the $p_{Ti}$ and $\cos\Theta$
distributions have significant changes,
especially for the massless visible particle case.
However, the proposed processes
in Eqs.~(\ref{eq:mssm})--(\ref{eq:mued})
are not affected since 
$\Gm_B/m_B$ is much smaller than 1\%.

\subsection{Longitudinal boost effect}
\label{subsec:long:boost}

\begin{figure}[!t]
\centering
  \includegraphics[scale=0.75]{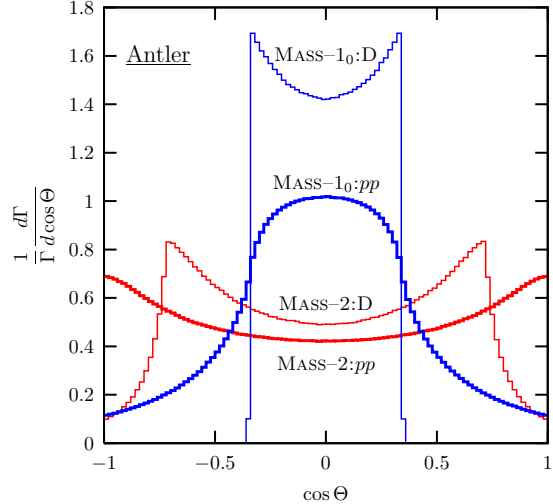}
  \caption[The $\cos\Theta$ distribution at the LHC for a UED model. ]
{\label{fig:longitudinal:boost}
Comparison of the $\cos\Theta$ distribution 
in the $D$ rest frame (thin curves) and in the $pp$ lab frame 
with $\sqrt{s}=14\ \tev$ (thick curves). 
 The mass parameters used here are
\textsc{Mass-1}$_0$ and \textsc{Mass-2} defined in 
Tables \ref{table:antler:mass:Z} and \ref{table:antler0:mass}.
}
\end{figure}

At hadron colliders,
the longitudinal motion of the particle $D$ is not determined
even when $D$ is singly produced.
Among the discussed kinematic variable,
only the $\cos\Th$ is affected,
which is defined with a momentum 
in the $D$ rest frame.
In order to see 
the longitudinal boost effects,
we convert the $\cos\Th$ distribution in the $D$ rest frame into
that in the $pp$ frame at the LHC, 
by convoluting with the parton distribution
functions of a proton. We have used
CTEQ6~\cite{cteq6}.
In Fig.\ref{fig:longitudinal:boost},
we compare the normalized 
$\cos\Theta$ distribution in the $D$-rest frame (thin curves) with that in
the $pp$ lab frame with $\sqrt{s}=14\ \tev$ (thick curves).
For simplicity we assume that the heavy particle $D$ 
is produced
through the $s$-channel gluon fusion and/or $q\bar{q}$ annihilation.

In the massive visible particle case (\textsc{mass-2}),
the cusped peaks vanish almost completely.
In the massless case (\textsc{mass-1}$_0$),
the pointed cusps become round,
very hard
to read.
We conclude that the cusp in the $\cos\Th$ distribution
is not observeble at the LHC.
In the $e^+ e^-$ collisions, however, 
the fixed c.m. energy removes the longitudinal boost ambiguity,
and thus the $\cos\Th$ cusp 
provides valuable information on the missing particle mass.

\subsection{Spin-correlation effect}
\label{subsec:spin:effect}

\begin{figure}[!t]
\centering
  \includegraphics[scale=0.75]{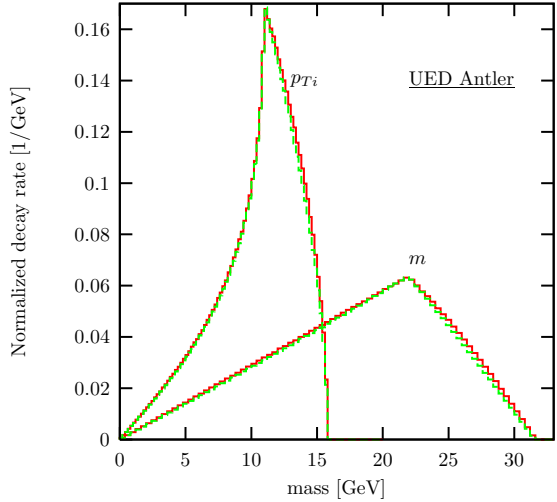}
  \caption[spin correlation]{\label{fig:spin:correlation}
The invariant mass distribution
$d\Gm/d m$ and the individual transverse momentum 
distribution $d\Gm/d p_{Ti}$
with and without spin correlations
of $\zt \to \lo \lo  \to \ell^- B^{(1)} \ell^+ B^{(1)}$  in the minimal UED model.
We have set $1/R =500\gev$ and $\Lm R=20$.
 }
\end{figure}

The effects of the spin-correlation by the full matrix elements
are different from new physics process to process.
In addition, if we consider the associated production of the particle $D$
in order to control the SM background,
the spin correlation effects
get intertwined with the additional $p_T$ and/or longitudinal boost effects.
To maximize the discovery significance, it is desirable to develop an 
individual strategy for each process in Eqs.(\ref{eq:mssm})--(\ref{eq:mued}),
which is beyond the scope of this paper.

Generically, 
the positions of cusps and endpoints are not affected by the spin
correlation effects since they are determined purely by the constrained 
phase space, {\it i.e.}, by the mass relations~\cite{Kim:2009si}. 
In order to see this feature,
we consider the $\zt$ decay in the framework 
of the minimal UED model (mUED)~\cite{ued}:
\bea
\label{eq:ztwo:mued}  
\zt \to \lo +\lo  \to \ell^- B^{(1)}+ \ell^+ B^{(1)} .
\eea

In Fig.~\ref{fig:spin:correlation},
we show the $m$ and $p_{Ti}$ distributions
including the full matrix elements of the process 
in Eq.~(\ref{eq:ztwo:mued}) at the LHC with $\sqrt{s}=14\tev$.
We have fixed $1/R =500\gev$ and $\Lm R =20$,
which generates the KK masses of
$m_D = 1048\gev$, 
$m_B =   515 \gev$, $m_a = 0$,
and $m_X =   500.9\gev$.
First finite width effects are negligible:
very degenerate mass spectrum in the mUED model
yields very small total decay width such that $\Gm_{B}/m_B \sim 10^{-4}$.
Second  the longitudinal boost effects do not apply to $m$ and $p_{Ti}$.
As shown in Fig.~\ref{fig:spin:correlation},
 the spin correlations
hardly change the $m$ and  $p_{Ti}$ distributions.
The distributions with and without the spin correlation effects are almost
identical.

Brief comments on the SM background
and detector simulation effects are in order
here.
In Ref.~\cite{cusp},
we have shown
that the cusp structure survives over
the SM backgrounds and the detector simulations
in a benchmark process
of $pp\to Z' \to \tilde{\ell}^+\tilde{\ell}^-
\to \ell^+ \neuo \ell^- \neuo$ in a supersymmetry model
with an extra $U(1)$ gauge field.
In addition the missing particle mass
as well as the intermediate particle mass
can be determined,
even though the uncertainty is about 10\%.
It was demonstrated that the analytic expression
for the invariant mass distribution 
is very helpful to reconstruct
the mass parameters by best-fitting.

\section{Summary and Conclusions}
\label{sec:conclusions}
In this paper, we have considered the antler decay topology 
of a parity-even heavy particle into two missing particles ($X_1$ and $X_2$)
and two visible particles ($a_1$ and $a_2$) via intermediate on-shell particles  ($B_1$ and $B_2$).
We studied the singularity structures in various kinematic distributions, especially  
non-smooth peaks called the cusps.
We show that the distributions of the invariant mass $m$ of $a_1$ and $ a_2$, 
the individual transverse momentum $p_{Ti}$, and the $\cos\Theta$ 
develop conspicuous cusp structures. 
We have provided the detailed derivations
for the positions of the cusps as well as the endpoints 
in terms of the particle masses.
The analytic functional forms of the invariant mass
and $\cos\Theta$ distributions  have been also given. 

The cusp and endpoint structures of the antler decay have a few advantages:
$(i)$ if the parent particle mass $m_D$ is known from other decay channels,
they can be used to determine
both the missing particle and intermediate particle masses;
$(ii)$ the cusped peaks are 
more identifiable than endpoints and kinks due to
higher statistics at the kinematical maxima; 
$(iii)$ the simple configuration of outgoing particles,
two visible particles and two missing particles,
avoids combinatoric complication, which is troublesome
in many missing particle mass measurement methods;
$(iv)$ 
the position of the cusp is independent of the 
$S$-matrix element such as the spin correlation effects, since it
is purely determined by the phase space.

We point out that the $p_{Ti}$ cusp and endpoint have some desirable features 
for observation.
The $p_{Ti}$ cusp tends to be sharp 
irrespective of mass parameter regions. It is complementary to 
the robust $m$ cusp, which  is sharp only when the masses are nearly degenerate.
The $p_{Ti}$ endpoint is always located
at fast-dropping end, which is easier to read off.
Finally, the cusp position for the massive visible particle case
is uniquely determined by the involved masses,
while the $m$ cusp has two-fold ambiguity.

It is noted that the cusp structures have some limitations for the missing 
particle mass determination, 
especially at the LHC.
The sharp cusped peaks in  the $\cos\Theta$ distribution
are not readily observable at the LHC,
due to the longitudinal boost of the produced $D$ particle.
The effects of the finite width 
of the intermediate particle could affect the cusp and endpoint
in the individual transverse momentum distribution.
However, for generically weakly coupled theories beyond the SM, the new particles for the antler decay
have relatively small decay widths, and thus the $p_{Ti}$ cusp is expected to be preserved.
The cusp in the invariant mass distribution,
which is the most robust observable at the LHC, 
is most pronounced for a degenerate mass spectrum.

In addition, the relations among different cusps and endpoints
help to identify the antler decay topology.
For example, the $m_T$ maximum is equal to
the $m$ maximum.
The cusp and endpoint of $p_{Ti}$ distribution
are half of those of $m$ distribution in the massless visible particle case.
One can use these facts for the consistency of the 
assumptions on the event topology.
Similar intriguing relations exist for the massive visible 
particle case.

In conclusion, 
if a new physics model accommodates an antler decay,
the measurement of kinematic cusps and endpoints can be helpful to 
determine the missing particle mass as well as the intermediate particle mass.
The proposed processes in various new physics models
are expected to have stable cusp and endpoint structures
in the $m$, $m_T$, and $p_{Ti}$ distributions at the LHC.

\begin{acknowledgments}
This work is supported in part by the U.S.
Department of Energy under grant No. DE-FG02-95ER40896, and in part by PITT PACC.
The work of JS was supported by WCU program
through the NRF funded by the MEST (R31-2008-000-10057-0).
\end{acknowledgments}

\appendix
\section{Kinematic distributions in the symmetric Antler decay}
\label{sec:invmassderivation}

In this Appendix, we
derive the kinematic distributions
of the invariant mass and the $\cos\Theta$ for the 
symmetric antler decays 
$D \to B_1 + B_2 \to a_1   X_1 + a_2  X_2$  
where $m_{B_1}\!= \!m_{B_2} \!\equiv\! m_B$, 
$m_{a_1} \!= \! m_{a_2} \!\equiv\! m_a$ and 
$m_{X_1}\!= \! m_{X_2} \!\equiv\! m_X$.
We first describe the four-body phase space and 
choose special 
internal phase variables.
For each four-momentum,
we specify the reference frame in the superscript.
For example, $k_1^{(B_1)}$ is the four-momentum of $a_1$ 
in the rest frame of $B_1$.
Since we calculate physical quantities mostly 
in the $D$ rest frame, 
we omit the superscript 
for the $D$ rest frame for simplicity.
Note that 
this is different from the notation in the main text 
where we omit the superscript for momenta in the lab frame.

\subsection{Four-body phase space}
\label{subsec:4phase:space}
We consider four body decays of
\bea
\label{eq:4body:decay}
D(P) \to a_1(k_1) +  a_2(k_2) +  X_1(k_3)+  X_2(k_4).
\eea
The differential decay rate of the process 
is
\beq
d \Gm = \frac{1}{2 m_D}\overline{|\M|^2} \;
d \Phi_4(P; k_1,k_2,k_3,k_4),
\eeq
where $\overline{|\M|^2} $ is the helicity amplitude squared,
and
$d \Phi_4$ is the element of four-body phase space, defined 
by~\cite{PDG,han:pheno}:
\begin{eqnarray}
d \Phi_4 ( P ; k_1, \dots, k_4 ) 
 = (2\pi)^4 \delta^4 
\left( P - \sum_{i=1}^4 k_i \right)\,
\prod_{i=1}^4 \frac{d^3 \mathbf{k}_i}{(2\pi)^3  2 E_i }.
\end{eqnarray}

If the decay in Eq.~(\ref{eq:4body:decay})
is through the antler decay, \textit{i.e.},
through $D \to B_1 B_2$
followed by $B_i\to a_i X_i (i=1,2)$,
the helicity amplitude squared $\overline{|\M|^2}$
has two propagator factors of $B_1$ and $B_2$.
Using the 
narrow width approximation $\Gamma_B/m_{B} \ll 1$, 
the matrix element squared can be expressed in terms of two Dirac delta functions:
\bea
\label{eq:Mhat:def}
\overline{|\M|^2}
& & \equiv
\widehat{|\M|^2}\; \frac{1}{(p_1^2-m_B^2)^2 + m_B^2 \Gm_B^2} \;
\frac{1}{(p_2^2-m_B^2)^2 + m_B^2 \Gm_B^2}
\nonumber \\
&  & \stackrel{\Gamma_B \ll m_{B}}{\longrightarrow}
\widehat{|\M|^2}\ 
\left(\frac{\pi}{m_{B }\Gm_B} \right)^{2}\ \delta(p_1^2-m_B^2)\ \delta(p_2^2-m_B^2),
\eea
where $p_1 = k_1 + k_3$ and $p_2 = k_2+k_4$
are the momentum of $B_1$ and $B_2$, respectively.
In this limit,
$\widehat{|\M|^2}$ does not develop any singular behavior and  
still remains as a smooth function containing spin correlation information. 

After the integration using delta functions,
the differential decay width 
is simplified to 
\beq
d \Gm = \frac{1}{2^{15}\pi^4m_D m_B^2\Gm_B^2}\,\widehat{|\M|^2}\,
\lm^{1/2}_B \
\lm_a \,d\cos\theta_1  d \cos\theta_2 d \phi,
\eeq
where $\lm_B =\lm\left(1,{m_B^2}/{m_D^2},{m_B^2}/{m_D^2} \right)$,
$\lm_a =\lm \left(1,{m_a^2}/{m_B^2},{m_X^2}/{m_B^2} \right)$,
and the standard  kinematic function is 
$\lm(a,b,c) = a^2+b^2+c^2-2 a b -2 a c + 2 b c$.
The polar angles of $\theta_1$ and $\theta_2$ 
and the azimuthal angle $\phi$ are 
defined in Fig.~\ref{fig:antphase}.
For simplicity, we use short-hand notations of
\bea
\label{eq:v1:v2:def}
v_1 \equiv \cos\theta_1, \quad v_2 \equiv \cos\theta_2,
\eea
and name $dv_1 dv_2 d \phi$ 
the normalized four-body phase space $\dphif$ 
of the antler decay:
\bea 
\label{eq:Phihat:def}
\dphif  = dv_1 dv_2 d \phi.
\eea

\subsection{Change of variables and 
the independence between angular variables}
\label{subsec:variable}
For a general two body decay of
$a \to b c$:
\bea
\label{eq:2body:decay}
a (p_a)\to b(p_b) + c(p_c),
\eea
the energy-momentum 
conservation in the rest frame of the parent particle $a$ 
leads to 
\begin{eqnarray}
m_a &=& E^{(a)}_b + E^{(a)}_c, \\
\no
{\bf p}^{(a)}_b &=& - {\bf p}^{(a)}_c. 
\end{eqnarray}
From the on-shell conditions of 
$p_i ^2 =m_i^2 ~(i=b,c)$, 
the energies and momenta of 
the particles $b$ and $c$ are simply expressed by
the rapidities $\eta_b$ and $\eta_c$:
\begin{eqnarray}
E^{(a)}_b &=& \frac{m_a^2 + m_b^2 - m_c^2 }{2 m_a} \equiv m_b \cosh \eta_b 
, \\
E^{(a)}_c &=& \frac{m_a^2 - m_b^2 + m_c^2 }{2 m_a} \equiv
 m_c \cosh \eta_c , \\
\left|{\bf p}^{(a)}_b\right| &=& \left|{\bf p}^{(a)}_c\right|
= m_b \sinh \eta_b = m_c \sinh \eta_c. 
\end{eqnarray}

For the symmetric antler decay,  
the same masses of $m_B \! \equiv\!  m_{B_1}\! =\!  m_{B_2}$ and
$m_a \! \equiv\!  m_{a_1}\!  =\! m_{a_2}$
lead to two independent rapidities: 
\begin{eqnarray}
\label{eq:eta:def}
\cosh \eta_B = \frac{m_D}{2 m_B}, \quad
\cosh \eta_a = \frac{m_B^2 - m_X^2 + m_a^2}{ 2 m_a m_B}. 
\end{eqnarray}

Now we present less intuitive but more convenient 
kinematic variables. First, we consider 
the rapidity of $a_2$
in the rest frame of $B_1$, not $B_2$, denoted by
$\alpha \equiv 
\eta^{(B_1)}_{a_2}$:
\begin{eqnarray}
\label{eq:alpha}
\cosh \alpha = \cosh 2\eta_B \, \cosh \eta_a 
- v_2 \, \sinh 2\eta_B \, \sinh \eta_a , 
\end{eqnarray}
where $v_2$ is defined in Eq.~(\ref{eq:v1:v2:def}).
The second useful variable is $u$,
the cosine of
the angle $\theta_{a_1 a_2}^{(B_1)}$ between $a_1$ and $a_2$ in the rest 
frame of $B_1$:
\begin{eqnarray}
\label{eq:u:v1v2cosphi}
u =
\frac{\mathbf{k}_1^{(B_1)} \cdot \mathbf{k}_2^{(B_1)}}
{\left|\mathbf{k}_1^{(B_1)} \right| \left| \mathbf{k}_2^{(B_1)}\right|}
=\frac{\sqrt{1-v_1^2}\sqrt{1-v_2^2}\, \cos\phi +
(\sinh 2\eta_B - v_2 \, \cosh 2\eta_B ) \, v_1 }{\cosh 2\eta_B - v_2 \, \sinh 2 
\eta_B}.
\end{eqnarray}
For simplicity, we define 
\bea
\label{eq:vp:vpp}
v^{\prime}_2 &=& \cosh 2\eta_B - v_2 \, \sinh 2\eta_B,
\\ \no
v^{\prime\prime}_2 &=& \sinh 2\eta_B - v_2 \, \cosh 2\eta_B.
\eea
Then the azimuthal angle  
$\phi$ is inversely obtained by 
\begin{eqnarray}
\label{eq:cos:phi}
\cos\phi =
\frac{u v_2' - v_1 v_2^{\prime\prime}}
{\sqrt{1-v_1^2}\sqrt{1-v_2^2}}. 
\end{eqnarray}

The advantage of this new angular variable $u$
is that $ d^2 \phif / d u d v_2 = \pi$:
$u$ and $v_2$ are independent variables
contrary to the expectation from the functional
dependence of $u$ on $v_1$ and $v_2$ 
in Eq.~(\ref{eq:u:v1v2cosphi}).
In order to show this non-trivial result,
we begin with $d^3 \phif / dv_1 dv_2 d\phi = 1$
in Eq.~(\ref{eq:Phihat:def}).
We change the variable $\phi$ into $u$ as
\begin{eqnarray}
\dphif=
d v_1 d v_2 d \phi &=& d v_1 d v_2 d u \left| \frac{\rd \phi}{\rd u} \right|
\\ \no
&=&d v_1 d v_2 d u 
\frac{v_2'}{\sqrt{(1-v_1^2)(1-v_2^2)-(u v_2'-v_1 v_2^{\prime\prime})^2}}
\\ \no
&\equiv&
d v_1 d v_2 d u \frac{v_2'}{\sqrt{f(u,v_1,v_2)}}.
\end{eqnarray}
Since the integrand $v^\prime_2 / \sqrt{f(u,v_1,v_2)}$ 
is not separable into products, 
$u$, $v_1$ and $v_2$ are not independent with one another. 
If we integrate one of 
the three variables, however,
we have the statistical independence of the remaining 
two variables. 
First $v_1$ and $v_2$ are independent variables by definition.
In order to see the independence of 
$v_2$ and $u$,
we integrate $v_1$ out for given $u$ and $v_2$. 
The integration limit of $v_1$ is matched with the roots of 
$f(u,v_1, v_2)=0$ for fixed $u$ and $v_2$. 
The result of the integration is a simple constant: 
\begin{eqnarray}
\int^{v_1^{(\rm max)}}_{v_1^{(\rm min)}}
d v_1 \frac{ v^\prime_2}{\sqrt{\left(v_1 - v_1^{(\rm min)}\right)
\left(v_1^{(\rm max)} - v_2\right) (v^\prime_2)^2 }} = \pi.
\end{eqnarray}   
Therefore $\dphif / du dv_2=\pi$ is also flat: 
$u$ and $v_2$ are independent.
Similarly, one can show the independence
of $u$ and $v_1$ from the symmetry under the exchange 
of $v_1$ and $v_2$.

\subsection{The invariant mass distribution}
\label{subsec:inv:mass}
The invariant mass $m$ of $a_1$ and $a_2$ is 
more simply expressed in terms of $\alpha$ 
and $u$ by
\begin{eqnarray}
\label{eq:Maa}
m^2 =  2 m_a^2 +  2 m_a^2 (\cosh \eta_{a} \cosh \alpha  
- u \sinh \eta_a \sinh \alpha),
\end{eqnarray}
where $\alpha$ and $u$ are defined in Eqs.~(\ref{eq:alpha}) and
(\ref{eq:u:v1v2cosphi}) respectively.
The expression in the parenthesis of Eq.~(\ref{eq:Maa})
is nothing but the
cosine hyperbolic of
the rapidity of the particle $a_1$
in the rest frame of $a_2$:
\begin{eqnarray}
\label{eq:chi:def}
\chi
&\equiv& \cosh \eta^{(a_1)}_{a_2}
 = \frac{m^2}{2 m_a^2} - 1 
=\cosh \eta_a \cosh \alpha  - u \sinh \eta_a \sinh \alpha.  
\end{eqnarray}

Now let us change variables 
from $(u,v_2)$ to $(\chi,\alpha)$:
\begin{eqnarray}
du dv_2 = d\chi d\alpha \frac{1}{\sinh 2\eta_B \sinh^2 \eta_a }. 
\end{eqnarray}
Note that the Jacobian factor is simply a constant.
From $d^2 \phif / du dv_2 = \pi$, we have
\begin{eqnarray}
\label{eq:dphi4:duv:pi}
\frac{\dphif}{d\chi} = 
\frac{\pi}{\sinh 2\eta_B \sinh^2 \eta_a}
\int^{\alpha_{\rm max} (\chi) }_{\alpha_{\rm min} (\chi)} 
d\alpha, 
\end{eqnarray}
where $\alpha_{\rm min} (\chi)$ and $\alpha_{\rm max} (\chi)$ are the minimum 
and maximum of $\alpha$ variable
at a given $\chi$, respectively. 

In order to obtain $\al_{\min}(\chi)$ and $\al_{\max}(\chi)$,
we use the conditions of 
$u\in [-1,1]$
and $v_2 \in [-1,1]$.
Then the definitions of $\cosh\alpha$ and $\chi$ in  Eqs.~(\ref{eq:alpha})
and (\ref{eq:chi:def}), respectively, constrain
the values of $\cosh\alpha$ and $\chi$ as
\bea
\label{eq:alpha:bound}
&&\cosh(2\eta_B-{\eta_a}) \leq \cosh\alpha \leq \cosh(2\eta_B+{\eta_a}),
\\ \label{eq:chi:bound}
&& \cosh(\al-{\eta_a}) \leq \chi \leq \cosh(\al+{\eta_a}).
\eea
Therefore,
the values of $\alpha_{\rm min} (\chi)$ and $\alpha_{\rm max} (\chi)$ in Eq.~(\ref{eq:dphi4:duv:pi})
depend on the relative size between 
$\eta_B$ and $\eta_a/2$ or $\eta_B$ and $\eta_a$.
This is related with the three different mass parameter
regions 
 of $\R_1$, $\R_2$, and $\R_3$ in Sec.~\ref{subsec:massive:m}:
 \beq
\R_1: \eta_B < \frac{{\eta_a}}{2},
\quad
\R_2: \frac{{\eta_a}}{2} < \eta_B <{\eta_a},
\quad
\mathcal{R}_3 :  {\eta_a}<\eta_B.
\eeq

\begin{figure}[!t]
\centering
  \includegraphics[scale=0.6]{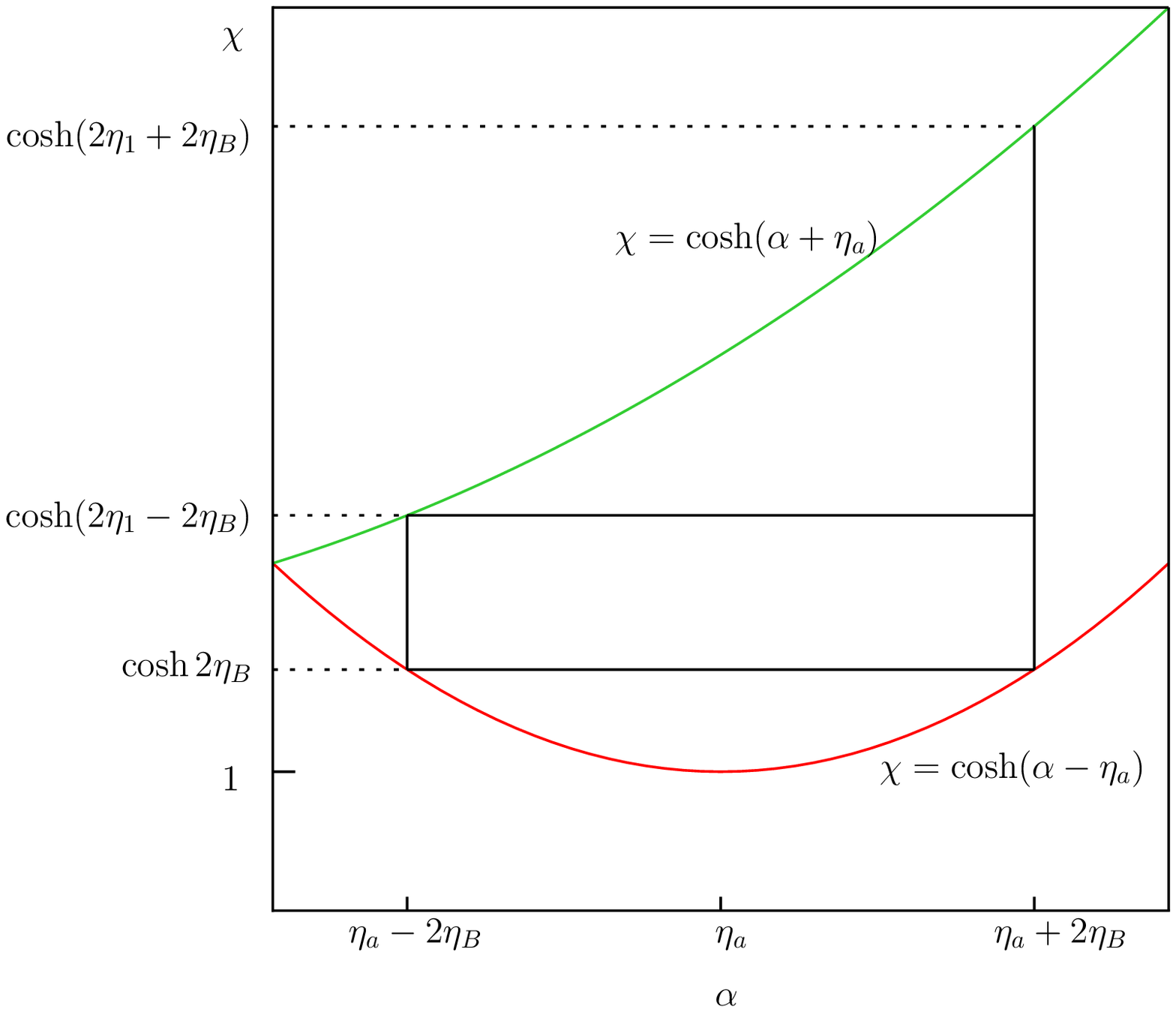}
  \caption{\label{fig:chialpha}
Allowed parameter space of $(\alpha,\chi)$ plane for the region $\R_1$.}
\end{figure}

Let us elaborate the derivation of
$\al_{\min}(\chi)$ and $\al_{\max}(\chi)$ for the region $\R_1$.
Figure \ref{fig:chialpha}
illustrates two curves of $\chi=\cosh(\alpha-\eta_a)$
and $\chi=\cosh(\alpha+\eta_a)$ in the parameter space of
$(\alpha,\chi)$.
Within the bound of $\eta_a - 2 \eta_B < \alpha
< \eta_a + 2 \eta_B$ as in Eq.(\ref{eq:alpha:bound}),
$\al_{\min}(\chi)$ and $\al_{\max}(\chi)$
are different according to the value of $\chi$,
summarized by 
\bea
\begin{array}{c|cc|c}
\hbox{ For }\R_1   & ~~~~~~\al_{\min}(\chi)~~~~~~ & ~~~~~~\al_{\max}(\chi)~~~~~~ & \int d \al\\
   \hline
  1<\chi<\chet & {\eta_a} - \achc & {\eta_a} + \achc & 2\achc\\
  \chet<\chi<c_{2\eta_a-2{\eta_B}} & {\eta_a}-2\eta_B & {\eta_a}+2\eta_B & 4\eta_B\\
  c_{2\eta_a-2{\eta_B}}<\chi<c_{2\eta_B+2{\eta_a}} & -{\eta_a}+\achc
  & {\eta_a}+2\eta_B & 2\eta_B+2{\eta_a}-\achc
\end{array}
\eea
Here we use the simplified notation of $c_x \equiv \cosh x$.
The derivations for $\R_2$ and $\R_3$ 
are similar and straightforward. 

With the help of Eq.~(\ref{eq:dphi4:duv:pi}), 
the final expressions for $d \phif/d m$ is given by
\bea
\left. \frac{1}{N} \frac{\dphif}{ d m}\right|_{\R_1}
&=&  
\left\{
   \begin{array}{ll}
   2m\cosh^{-1}\left(\frac{m^2}{2 m_a^2}-1\right) ,
               & \hbox{ if } 1<\chi<c_{ 2\eta_B}; \\
   4 \eta_B m ,
               & \hbox{ if } c_{ 2 \eta_B} <\chi <
c_{2({\eta_a}-\eta_B)}; \\
   2(\eta_a+\eta_B)m -m\cosh^{-1}\left(\frac{m^2}{2 m_a^2}-1\right)  ,
               & \hbox{ if } c_{ 2({\eta_a}-\eta_B)} < \chi <
c_{ 2({\eta_a}+\eta_B)},
   \end{array}
 \right.
\eea
\bea
\!\!\!\!\left. \frac{1}{N}\frac{\dphif}{ d m} \right|_{\R_2}
&=& 
\left\{
   \begin{array}{ll}
   2m\cosh^{-1}\left(\frac{m^2}{2 m_a^2}-1\right) ,
               & \hbox{ if } 1<\chi<c_{ 2(\eta_a-\eta_B)}; \\
   2(\eta_a-\eta_B)m +m\cosh^{-1}\left(\frac{m^2}{2 m_a^2}-1\right)   ,
               & \hbox{ if } c_{2({\eta_a}-\eta_B)}<\chi<
c_{ 2\eta_B}; \\
   2(\eta_a+\eta_B)m -m\cosh^{-1}\left(\frac{m^2}{2 m_a^2}-1\right)  ,
               & \hbox{ if } c_{ 2\eta_B} <\chi <
c_{ 2({\eta_a}+\eta_B)},
   \end{array}
 \right.
\eea
\bea
\left. \frac{1}{N} \frac{\dphif}{d m} \right|_{\R_3}
&=&  
\left\{
   \begin{array}{ll}
   2(\eta_a-\eta_B)m +m\cosh^{-1}\left(\frac{m^2}{2 m_a^2}-1\right),
               & \hbox{ if } c_{2({\eta_a}-\eta_B)}<\chi
< c_{2\eta_B}; \\
   2(\eta_a+\eta_B)m -m\cosh^{-1}\left(\frac{m^2}{2 m_a^2}-1\right),
               & \hbox{ if } c_{ 2\eta_B} < \chi 
< c_{ 2({\eta_a}+\eta_B)}, 
   \end{array}
 \right. 
\eea
where the normalization factor $N$ is
\begin{eqnarray}
N = \frac{\pi}{\sinh {2\eta_B}} \frac{1}{(m_a \sinh \eta_a)^2}.
\end{eqnarray} 

\subsection{The angular distribution $d\Gamma / d\cos\Theta$}
\label{appendix:cosTh}
In this subsection, we derive $\dphif/d \cos\Th$,
restricting ourselves to 
the massless visible 
particle case ($m_{a_1}\!=\!m_{a_2}\!=\!0$).
Recall that $\Theta$ is the angle of a visible particle, 
say $a_1$, in the c.m.~frame of $a_1$ and $a_2$, with respect to 
their c.m.~moving direction in the $D$ rest frame.
For $\dphif/d \cos\Th$,
we begin with 
$d^3 \widehat{\Phi}_4/d v_1 dv_2 d \phi =1$.
The key point is the Jacobian factor from
$d v_1 dv_2 d \phi$   to $d \cos\Th$.
For this goal,
we first obtain the analytic expression of $\phi$ in terms of
$v_1$, $v_2$, and $\cos\Th$.

In the case of $m_a=0$,
the $k_1$ and $k_2$ four-momenta in the $B_1$ rest frame
become
\bea
k_{1}^{(B_1)} &=& E^{(B_1)}_1 (1, \sqrt{1-v_1^2},0,-v_1)
= E_\ell  ( 1,\hat{\mathbf{k}}_1),
\\ \no
k_2^{(B_1)} &=& E^{(B_1)}_1  (v_2^{\prime}, \sqrt{1-v_2^2}\cos\phi,\sqrt{1-v_2^2}\sin\phi,
-v_2^{\prime\prime})
= E_\ell  v_2' ( 1,\hat{\mathbf{k}}_2)
,
\eea
where $E_\ell = m_B(1-m_X^2/m_B^2)/2$,
$\hat{\mathbf{k}}_i = \mathbf{k}_i/|\mathbf{k}_i| (i=1,2)$,
and the definitions of
$v_2'$ and $v_2^{\prime\prime}$ are in Eq.~(\ref{eq:vp:vpp}).
Defining $k\equiv k_1 + k_2 = (E_{\rm cm}, \mathbf{k})$,
we have some useful expressions of
\bea
\label{eq:Maa:uv2}
m^2 &=& 2 E_\ell^2 v_2' (1 - u),
\\ \no
E_{\rm cm}^{(B_1)} &=&  E_\ell (1+v_2'),
\\ \no
|\vec{k}^{(B_1)}|^2 &=& 
E_\ell^2 \{ 1+ 2 v_2' u + (v_2')^2 \}.
\eea

Now the Lorentz transformation matrix 
from the $B_1$ rest frame to
the c.m. frame of $a_1 a_2$ is
\bea
\Lm^{(a_1a_2\leftarrow B_1)} =
\left(
  \begin{array}{cc}
    \gm_{\rm cm } & -\dfrac{ \mathbf{k}^{(B_1)\, T} }{m} \\[0.4cm]
  - \dfrac{ \mathbf{k}^{(B_1)}}{m} &~~~~~ I_{3\times 3} +
\left( \gm_{\rm cm }-1\right) \hat{\mathbf{k}}^{(B_1)} 
\hat{\mathbf{k}}^{(B_1)\,T }  \\
  \end{array}
\right),
\eea
where $\gm_{\rm cm } ={E_{\rm cm }^{(B_1)}}/{m}$, 
and the superscript $T$ denotes the transpose of the vector.
The three-momentum of $a_1$ particle in the  c.m.~frame of $a_1$ and $ a_2$
is
\bea
\mathbf{k}_1^{(\rm cm )} =
\left\{
- \frac{E_{\rm cm}^{(B_1)}}{m}\mathbf{k}^{(B_1)} + \mathbf{k}_1
+\left( \gm_{\rm cm }-1 \right)
(\hat{\mathbf{k}}^{(B_1)} \cdot \mathbf{k}_1^{(B_1)}) \hat{\mathbf{k}}^{(B_1)}
\right\}
.
\eea
Since $k$ in the $D$ rest frame is $k^{(D)} = - P_D^{(\rm cm )}
=-\Lm^{(a_1a_2\leftarrow B_1)} P_D^{(B_1)}$, we have
\bea
\mathbf{k}^{(D)}= -
\left\{
- \frac{E_{D}^{(B_1)}}{m} \mathbf{k} + \mathbf{P}_D^{(B_1)} +
\left( \gm_{\rm cm } -1 \right)
(\hat{\mathbf{k}}^{(B_1)} \cdot \mathbf{P}_D^{(B_1)})\hat{\mathbf{k}}^{(B_1)}
\right\}
.
\eea

The dot-product of $\mathbf{k}_1^{(\rm cm )}$
and $ \mathbf{k}^{(D)}$ leads to $\cos\Th$:
\bea
\cos\Theta
 = \frac{\mathbf{k}_1^{(\rm cm )} \cdot \mathbf{k}^{(D)}}{ |\mathbf{k}_1^{(\rm cm )} | |\mathbf{k}^{(D)}| }
\,.
\eea
Finally we express $\cos\Th$ in terms of $(v_1,v_2,\phi)$:
\bea
\label{eq:cosTh:v1v2cosphi}
\cos\Th
&=& \frac{(v_2-v_1) \, s_{\eta_B}}{\sqrt{2 - \frac{1}{2} (v_1+v_2)^2 
+ \frac{1}{2} (v_1-v_2)^2 \, c_{ 2\eta_B}  + 2
\sqrt{(1-v_1^2)(1-v_2^2)}\, \cos\phi}}.
\eea
Note that the maximum of $\cos\Th$ occurs when
$v_1 = \pm 1 $ and $v_2 = \mp 1$, \textit{i.e.,}
when the visible particles $a_1$ and $a_2$ are moving in the same direction.
The maximum of $\cos\Th$ in the $D$-rest frame is then
\bea
\label{eq:cosTh:max}
\left|\cos\Th \right|_{\rm max} = \tanh\eta_B.
\eea

Finally $\cos\phi$ is expressed
in terms of $v_1$, $v_2$, and $\Theta$:
\bea
\cos\phi =
\dfrac{-1 + \frac{1}{4} (v_1+v_2)^2 +
\frac{1}{4} 
\left(\frac{2s_{\eta_B}^2}{\cos^2\Th}- c_{2\eta_B}
\right) (v_1-v_2)^2}{\sqrt{(1-v_1^2)(1-v_2^2)}}
\,,
\eea
where $s_\eta \equiv \sinh \eta$ for simplicity.
For the Jacobian factor, 
we introduce three independent variables, $v_+$, $v_-$, and $t$, defined by
\bea
v_\pm = v_1 \pm v_2,
\quad
t = \frac{2s_{\eta_B}^2}{\cos^2\Th}- c_{2\eta_B}.
\eea
Note that  the maximum of $|\cos\Th|$
in Eq.(\ref{eq:cosTh:max}) leads to the $t$ integration range as
$1 \leq t< \infty$.
Since
\bea
\dphif =
d v_1 d v_2 d \phi = \frac{1}{2}d v_+ dv_- d t
\left|
\frac{\rd \phi}{\rd t}
\right|
,
\eea
the differential four-body phase space 
with respect to $t$ is
\bea
\frac{\dphif }{d t}
= \frac{1}{4} \int d v_+ d v_-^2
\frac{1}{ \sqrt{v_-^2 (1-t^2) 
+  8 t -2 v_+^2 t -8 -2 v_+^2}}.
\eea
The integration range is
\bea
\label{eq:v-:range} &&
0\leq v_-^2 \leq 
\frac{-8-2 v_+^2 + 8 t - 2 v_+^2t }
{t^2-1},
\\ \label{eq:v+:range} &&
0 \leq v_+^2 \leq \frac{4(t-1)}{t+1}.
\eea
Finally the integration over $v_-$ and $v_+$ yields
\bea
\frac{\dphif}{d \cos\Th}
= 4 \sqrt{2} \pi \sinh^2 \eta_B \frac{1}{\sin^3\Th}.
\eea

\section{The invariant mass distribution of generic antler decays}
\label{sec:general}
In this section, we present the analytic expression of the invariant 
mass distribution of generic non-symmetric antler decays with
 $m_{B_1} \!\neq\! m_{B_2}$, $m_{a_1} \!\neq\!  m_{a_2}$ and 
$m_{X_1} \!\neq\! m_{X_2}$. 
The derivation is very similar to Appendix 
\ref{sec:invmassderivation}, 
but in this general case the mass parameter space is divided into finer twelve regions. 
Since the derivation of the formulae for each region 
is long and tedious, we show 
only the results here.  

\subsection{Massive visible particles ($m_a \neq 0$)}
\label{subsec:inv:mass:massive}

In generic antler decays, 
there are in general six different rapidity parameters, 
given by 
\begin{eqnarray}
\cosh \eta_{X_1} &=& 
\frac{m_{B_1}^2 + m_{X_1}^2 - m_{a_1}^2}{2 m_{X_1} m_{B_1} }, \qquad
\cosh \eta_{X_2} =
\frac{m_{B_2}^2 + m_{X_2}^2 - m_{a_2}^2 }{2 m_{X_2} m_{B_2}}, \nonumber \\
\cosh \eta_{B_1} &=&
\frac{m_D^2 + m_{B_1}^2 - m_{B_2}^2 }{2 m_{B_1} m_D}, \qquad
\cosh \eta_{B_2} = 
\frac{m_D^2+ m_{B_2}^2 - m_{B_1}^2}{2 m_{B_2} m_D}, \nonumber \\
\cosh \eta_{a_1} &=&
\frac{m_{B_1}^2 + m_{a_1}^2 - m_{X_1}^2}{2 m_{a_1} m_{B_1}}, \qquad
\cosh \eta_{a_2} =
\frac{m_{B_2}^2 + m_{a_2}^2 - m_{X_2}^2}{2 m_{a_2} m_{B_2}}.
\end{eqnarray}

We define 
\begin{eqnarray}
\eta_{++} &=& \eta_{B_1} + \eta_{B_2} + \eta_{a_1} + \eta_{a_2}, \\
\eta_{+-} &=& |\eta_{B_1}+\eta_{B_2}+\eta_{a_1}-\eta_{a_2}|,  \\
\eta_{-+} &=& |\eta_{B_1}+\eta_{B_2}-\eta_{a_1}+\eta_{a_2}|, \\
\eta_{--} &=& |\eta_{B_1}+\eta_{B_1}-\eta_{a_1}-\eta_{a_2}|.  
\end{eqnarray}
From positive definite definition of the rapidity,
$\eta_{++}$ is the larges among four $\eta_{\pm\pm}$'s.
However the relative size of the other three $\eta$'s
is different according to the mass parameters.
We order $\eta_{+-}$, $\eta_{-+}$ and $\eta_{--}$ 
and name them to be
$\eta_1 \leq \eta_2 \leq \eta_3$. 
We have 6 regions depending on this
ordering:
\begin{eqnarray}
&& \eta_{+-} \leq \eta_{-+} \leq \eta_{--}, \qquad 
\eta_{-+} \leq \eta_{+-} \leq \eta_{--}, \nonumber \\
&& \eta_{+-} \leq \eta_{--} \leq \eta_{-+}, \qquad
\eta_{-+} \leq \eta_{--} \leq \eta_{+-}, \nonumber \\
&& \eta_{--} \leq \eta_{+-} \leq \eta_{-+}, \qquad
\eta_{--} \leq \eta_{-+} \leq \eta_{+-}. \nonumber
\end{eqnarray}

To obtain $d \Gm/d m$,
we introduce the general $\chi$, defined by
\bea
\chi \equiv \cosh \eta_{a_2}^{(a_1)} 
= \frac{m^2 - m_{a_1}^2 - m_{a_2}^2}{2 m_{a_1} m_{a_2}}.
\eea
The general invariant mass distribution have
12 different cases in total, given by
\begin{itemize}
\item If $|\eta_{B_1} + \eta_{B_2} - \eta_{a_2}| \geq \eta_{a_1}$ or 
$\eta_{B_1} + \eta_{B_2} + \eta_{a_2} \leq \eta_{a_1}$, 
\begin{eqnarray}
\frac{1}{\tilde{N}}\frac{\dphif}{  dm} =  
\begin{cases}
\begin{array}{ll}
-\eta_1 m + m \cosh^{-1} 
\left( \frac{ m^2 - m_{a_1}^2 - m_{a_2}^2 }{2 m_{a_1} m_{a_2} }\right), &
\mbox{ if $c_{\eta_1}  \leq \chi \leq c_{\eta_2}$, } \\
\eta_2 -\eta_1,  &
\mbox{ if $c_{\eta_2} \leq \chi \leq c_{ \eta_3}$, } \\
\eta_{++} - \cosh^{-1} 
\left( \frac{ m^2 - m_{a_1}^2 - m_{a_2}^2 }{2 m_{a_1} m_{a_2} }\right), &
\mbox{ if $c_{ \eta_3} \leq \chi \leq c_{ \eta_{++}}$, } \\
0, & \mbox{ otherwise.}
\end{array}
\end{cases}
\end{eqnarray}
\item If  $|\eta_{B_1}+\eta_{B_2}-\eta_{a_2}| < \eta_{a_1} 
< \eta_{B_1}+\eta_{B_2}+\eta_{a_2}$, 
\begin{eqnarray}
\frac{1}{\tilde{N}}\frac{\dphif}{dm} =  
\begin{cases}
\begin{array}{ll}
2 m \cosh^{-1} 
\left( \frac{ m^2 - m_{a_1}^2 - m_{a_2}^2 }{2 m_{a_1} m_{a_2} }\right), & 
\mbox{ if $1 \leq \chi \leq c_{\eta_1}$, } \\
-\eta_1 m + m\cosh^{-1} 
\left( \frac{ m^2 - m_{a_1}^2 - m_{a_2}^2 }{2 m_{a_1} m_{a_2} }\right), &
\mbox{ if $ c_{\eta_1} \leq \chi \leq c_{\eta_2}$, } \\
(\eta_1 + \eta_2)m,  &
\mbox{ if $c_{\eta_2} \leq \chi \leq c_{ \eta_3}$, } \\
\eta_{++}m - m\cosh^{-1} 
\left( \frac{ m^2 - m_{a_1}^2 - m_{a_2}^2 }{2 m_{a_1} m_{a_2} }\right), &
\mbox{ if $c_{\eta_3} \leq \chi \leq c_{ \eta_{++}}$, } \\
0, & \mbox{ otherwise.}
\end{array}
\end{cases}
\end{eqnarray}
\end{itemize} 
Here the normalization factor $\tilde{N}$ is given by 
\begin{eqnarray}
\tilde{N} = \frac{\pi}{m_{a_1} m_{a_2} \sinh 2\eta_B \sinh \eta_{a_1} 
\sinh \eta_{a_2}}.
\end{eqnarray}
Note that the minimum of the invariant mass distribution can be different from 
$m_{a_1}+m_{a_2}$,
according to the mass 
parameter regions.
Crucial is whether the kinematic configuration
that $a_1$ and $a_2$ are 
relatively at rest is allowed.

\subsection{Massless visible particles ($m_a =0$)}
\label{subsec:inv:mass:massless}
In this subsection,
we present the invariant mass distribution 
for massless visible particle but different intermediate particle cases,
\textit{i.e.}, when
$m_{B_1} \neq m_{B_2}$ and $m_{a_1}=m_{a_2}=0$.
In this case, $\eta_{--}$ is always larger than
$\eta_{+-}$ and 
$\eta_{-+}$, leading to $\eta_3 = \eta_{--}$.
Here we need to consider only the leading terms 
of ${\mathcal O}\left(m_{a_1}^{-1} m_{a_2}^{-1} \right)$,
which are absent in 
$\cosh\eta_{+-}$ 
and $\cosh \eta_{-+}$.
Therefore, the invariant
mass distribution is divided into three regions. 
Using $\cosh^{-1} x =\ln(x+\sqrt{x^2-1}) 
\approx \ln (2x)$ for $x \gg 1$, we have
\begin{eqnarray}
\frac{\dphif}{dm}\propto 
\begin{cases}
\begin{array}{ll}
m \log \left( \frac{m_{\rm max}^{(0)}}{m_{\rm cusp}^{(0)}} \right), & 
\mbox{ if $ 0 < m < m_{\rm cusp}^{(0)}$ ;  } \\
m \log \left( 
\frac{m_{\rm max}^{(0)}}{m}
\right),
& 
\mbox{ if $ m_{\rm cusp}^{(0)} < m  < m_{\rm max}^{(0)}$;} \\
0, & \mbox{ otherwise, } 
\end{array}
\end{cases}
\end{eqnarray}
where 
\begin{eqnarray}
m_{\rm cusp}^{(0)} &=& 
\sqrt{\left(\frac{m_{B_1}^2 - m_{X_1}^2}{m_{B_1}}\right)
\left(\frac{m_{B_2}^2 - m_{X_2}^2}{m_{B_2}}\right) }  \,
\exp \left(-\frac{\eta_{B_1}+\eta_{B_2}}{2}\right), 
\\
m_{\rm max}^{(0)} &=& 
\sqrt{
\left(\frac{m_{B_1}^2 - m_{X_1}^2}{m_{B_1}}\right)
\left(\frac{m_{B_2}^2 - m_{X_2}^2}{m_{B_2}}\right) } \, \exp \left(\frac{\eta_{B_1}+\eta_{B_2}}{2}\right).
\end{eqnarray}
This is the generalized results of Eqs.~(\ref{eq:masscusp0})
and (\ref{eq:massmax0}). 
Note that the product 
$m_{\rm cusp}^{(0)}m_{\rm max}^{(0)}$ depends only on the
second step decays of 
$B_1 \rightarrow X_1 a_1$ and $B_2 \rightarrow X_2 a_2$
while the ratio 
$m_{\rm max}^{(0)}/m_{\rm cusp}^{(0)}$ only on the first step
decay of $D \rightarrow B_1 B_2$.

\end{document}